\documentclass[superscriptaddress, prd, aps,amsmath,amssymb,showpacs,showkeys, onecolumn]{revtex4-2}
\usepackage[dvips]{graphicx,color}
\usepackage{subfig}
\usepackage{times}
\usepackage{xcolor}
\usepackage[%
  colorlinks=true,
  urlcolor=blue,
  linkcolor=red,
  citecolor=blue
]{hyperref}
\usepackage{orcidlink}

\begin{document}
\title{Joule-Thomson effect and Efficiency of deformed AdS-Schwarzschild black hole in presence of quintessence}

\author{Dhruba Jyoti Gogoi\orcidlink{0000-0002-4776-8506}}
\email[Email: ]{moloydhruba@yahoo.in}
\affiliation{Department of Physics, Madhabdev University, Narayanpur, Lakhimpur 784164, Assam, India}
\affiliation{Research Center of Astrophysics and Cosmology, Khazar University, Baku, AZ1096, 41 Mehseti Street, Azerbaijan}

\author{Ronit Karmakar\orcidlink{0000-0002-9531-7435}}
\email[Email: ]{ronit.karmakar622@gmail.com}
\affiliation{Department of Physics, Moran College, Moranhat, Charaideo 785670, Assam, India}

\author{Jyatsnasree Bora\orcidlink{0000-0001-9751-5614}}
\email[Email: ]{jyatnasree.borah@gmail.com}
\affiliation{Department of Physics, Madhabdev University, Narayanpur, Lakhimpur 784164, Assam, India}
\affiliation{Pacif Institute of Cosmology and Selfology (PICS) Sagara, Sambalpur 768224, Odisha, India}

\author{Pohar Buragohain \orcidlink{0009-0004-5223-5618}}
\email[Email: ]{buragohainpohar@gmail.com}
\affiliation{Department of Physics, Moran College, Moranhat, Charaideo 785670, Assam, India}

\author{Chandika Gogoi \orcidlink{0009-0002-4820-0717}}
\email[Email: ]{gogoichandika63@gmail.com}
\affiliation{Department of Physics, Moran College, Moranhat, Charaideo 785670, Assam, India}

%\date{}
\begin{abstract}

We study the Joule–Thomson expansion and extended thermodynamics of a modified black hole characterised by the parameters $\alpha$, $\beta$, and $\sigma$. Analysis of the Hawking temperature, Joule–Thomson coefficient, inversion curves, and isenthalpic trajectories shows that these parameters significantly modify the heating–cooling behaviour and thermal stability of the system. The deformation parameter $\alpha$ and control parameter $\beta$ shift the temperature minimum, enlarge the cooling region, and raise the inversion temperature, while $\sigma$ produces a weaker but consistent influence. The heat-engine analysis reveals that $\alpha$ enhances efficiency, whereas higher $\beta$ and $\sigma$ reduce it. Overall, the results demonstrate that geometric deformation and quintessence jointly govern the unified thermodynamic structure of the black hole.

\end{abstract}

%\pacs{04.30.Tv, 04.50.Kd}

\maketitle
\section{Introduction}\label{sec01}

Black holes constitute regions of spacetime where gravitational attraction is sufficiently intense to prevent the escape of particles and electromagnetic radiation. At their centre, both density and curvature diverge, giving rise to a singularity. The process of crossing the event horizon is therefore irreversible \cite{Witten:2024upt}. More than a century ago, in 1915, Einstein formulated General Relativity (GR) \cite{Einstein:1915ca}. Although extraordinarily successful, GR exhibits well-known limitations: it predicts physical singularities such as those at the centres of black holes where the theory ceases to remain valid, it fails to reconcile with quantum mechanics, and it relies on unobserved components like Dark Matter (DM) and Dark Energy (DE) \cite{Sotiriou:2008rp}. Consequently, despite its remarkable empirical successes, GR is not regarded as the ultimate or complete description of gravitation \cite{Sotiriou:2008rp}. 

A transformative development in gravitational physics is the realisation that black holes behave as thermodynamic systems, fundamentally altering our understanding of GR and its interface with Quantum Field Theory (QFT) \cite{Hansen:2016ayo, Cai:2014znn, Carlip:2014pma, Soroushfar:2025yhr, Karmakar:2024xwr, Gogoi:2025rcn, Gogoi:2025ied}. Classical black holes, characterised solely by mass, charge, and angular momentum, obey dynamical laws that closely mirror the four laws of thermodynamics. This correspondence gained physical significance when Bekenstein proposed that black holes possess entropy, motivated by the paradox that ordinary entropy appears to be lost when matter falls into a black hole \cite{Bekenstein:1972tm}. To preserve the second law of thermodynamics, he introduced the notion that black hole entropy is proportional to the surface area of the event horizon \cite{Bekenstein:1972tm}. This led to the formulation of the Generalised Second Law (GSL), which asserts that the sum of ordinary entropy and black hole entropy never decreases, thereby resolving the apparent entropy loss paradox. Black holes also exhibit negative specific heat, preventing them from attaining stable thermal equilibrium with an arbitrarily large heat bath and complicating attempts to employ standard statistical-mechanical ensembles in gravitational contexts \cite{Bora:2025zja}. Recently, the topology of black hole thermodynamics has also received some attention from researchers \cite{Fang:2022rsb, Wei:2022dzw, Yerra:2022eov, Wei:2021vdx, Gogoi:2025ied}. 

The Joule–Thomson (J–T) expansion provides an important thermodynamic framework to investigate a thermal system as well as black hole systems in different modified theories of gravity. In this process, a gas initially at high pressure expands through a porous barrier into a region of lower pressure while maintaining constant enthalpy. For black holes, enthalpy corresponds to the mass \cite{Okcu:2016tgt, Gogoi:2024ypn, Sekhmani:2024jli, Gogoi:2025rcn}. The J–T expansion allows one to examine heating–cooling behaviour and derive inversion temperatures. The temperature variation with respect to pressure at constant enthalpy is quantified by the J–T coefficient,
$\mu = \left( \frac{\partial T}{\partial P} \right)_H$,
whose sign determines whether the system heats or cools during expansion. Although pressure variation is always negative during expansion, the temperature variation may assume either sign depending on the characteristics of the system. A positive temperature change implies a negative J–T coefficient, indicating heating, whereas a negative temperature change associated with a positive J-T coefficient signals cooling. At the inversion temperature $T_i$, the coefficient $\mu$ vanishes. The corresponding pressure is the inversion pressure $(P_i)$, and the point $(T_i, P_i)$ marks the transition between heating and cooling regimes. Regions above the inversion curve correspond to cooling, whereas regions below correspond to heating \cite{Okcu:2016tgt}.

Extensive studies have established that the temperature–pressure (T–P) diagram provides a robust diagnostic tool for distinguishing heating and cooling regions \cite{Kruglov:2023zci, Tataryn:2021liq, Dehghani:2023yph, Kruglov:2023ogn, Alipour:2024xxs, Wang:2024jtp, Kruglov:2023ktg, Wang:2024jlj, Bakhtiarizadeh:2025wbd, Fatima:2025aqb, Dai:2025fxl, Wang:2025alf, Media:2025gvl}. In a detailed treatment of J–T expansion for black holes in Einstein gravity was presented by \"Okc\"u and Aydıner \cite{Okcu:2016tgt}, who analysed the J–T effect for charged AdS black holes and identified both parallels and departures from Van der Waals fluids. Their subsequent work extended the analysis to Kerr–AdS black holes within the extended phase space \cite{Okcu:2017qgo}, effectively characterising cooling and heating regions from the T–P diagram. Later studies generalised these ideas to quintessence holographic superfluids of Reissner–Nordström black holes in $f(R)$ gravity \cite{Ghaffarnejad:2018exz, Chabab:2018zix}. The influence of spacetime dimensionality on J–T expansion was examined in Ref. \cite{Chabab:2018zix}. Furthermore, the J–T expansion of charged accelerating AdS black holes in $f(R)$ gravity demonstrated that the inversion curve consistently corresponds to the lower branch, implying that the black hole remains in a cooling regime throughout expansion \cite{Rostami:2019ivr}. Quantum-corrected Schwarzschild–AdS black holes surrounded by quintessence have also been analysed to explore their thermodynamic and shadow properties \cite{Hamil:2023zeb}.

Recent investigations of black holes in metric-affine gravity reveal that inversion temperature and pressure increase with the strength of metric-affine corrections, while they decrease with the charge $(q_e)$ \cite{Yasir:2023qho}. Higher-dimensional charged dilatonic black holes with potential $V(\phi)$ have similarly been studied, highlighting the influence of dimensionality and dilaton parameters on J–T behaviour \cite{Zhang:2022cdp}. Studies of charged AdS black holes surrounded by modified Chaplygin gas show that the divergence point and zero point of the J–T coefficient shift rightward as charge increases \cite{Zhang:2024fxj}. {Another work \cite{Ali:2024jqo} discusses the Joule-Thomson expansion and thermodynamics of a non-linear charged AdS black hole solution, and analyses the effect of model parameters on the inversion curves. Ref. \cite{Chaudhary:2022sfg} uses a simplified framework for thermodynamic geometry of AdS black holes in Einstein Maxwell scalar theory, constructing consistent geometric descriptions across phase spaces and examining stability under thermal fluctuations, extending the analysis to regularized Lovelock theory.} These works collectively inspired a substantial body of literature on J–T expansion across diverse black hole geometries \cite{new3,new4,new5,new6,Kruglov:2023zci, Tataryn:2021liq, Dehghani:2023yph, Gogoi:2023ntt, Paul:2023mlh, Kruglov:2023ogn, Alipour:2024xxs, Wang:2024jtp, Kruglov:2023ktg, Wang:2024jlj, Ladghami:2024yjn, Ahmed:2025qza, Bakhtiarizadeh:2025wbd, Fatima:2025aqb, Dai:2025fxl, Javed:2025dit, Waseem:2025bwb, Wang:2025alf, Media:2025gvl, Liu:2025ulr, Javed:2024ohv, Liu:2024zzr, Mustafa:2024fau, Javed:2024nnt, Rezaei:2024qpd, Yasir:2024gza, Masmar:2023qol, Zhang:2022cdp, Sekhmani:2022hir, Feng:2022wnq, Barrientos:2022uit, Kruglov:2022lnc, Xing:2021gpn, Meng:2021cgb, Graca:2021ker, Biswas:2021uop, Zhang:2021kha, Yin:2021akt, MoraisGraca:2021ife, Liang:2021xny, Rajani:2020mdw, Lan:2019kak, Rostami:2019ivr, Nam:2019zyk, MahdavianYekta:2019dwf, RizwanCL:2018cyb, Mo:2018rgq, Paul:2024yrc}.

Parallel to these developments, black holes have also been conceptualised as working substances in thermodynamic heat engines. Holographic black hole heat engines, capable of performing mechanical work, have been analysed extensively in \cite{Johnson:2018amj}. Rotating black hole heat engines extract mechanical work from the rotational energy stored in the ergosphere \cite{Penrose:1969pc}. Additional analyses of efficiency and work extraction for a variety of black hole heat engine models appear in \cite{Setare:2015xaa, Johnson:2015fva, Hendi:2017bys, Mo:2017nhw, Nam:2019zyk, Roy:2021ucl, Nag:2023ibb, Roy:2023qqy, Fatima:2025pny}. Of particular note, Ref. \cite{Rajani:2019ovp} examined the efficiency and Carnot efficiency of Bardeen–AdS black hole heat engines and demonstrated that quintessence fields enhance thermodynamic efficiency.

With this brief review of the relevant literature, the present study investigates the thermodynamic properties, J-T expansion, and heat engine efficiency of deformed AdS-Schwarzschild black holes in a quintessence background. { As mentioned earlier, constructing regular, non-singular black hole models is a critical step toward a complete gravitational theory. In this context, the deformed black hole geometry investigated in this work is not proposed as the final, fundamental theory of quantum gravity, but rather as a highly effective and necessary phenomenological testbed. By introducing a deformation parameter $\alpha$ to ensure the rapid asymptotic decay of the central energy density, and a control parameter $\beta$ to encode nonlinear regularization effects, this model captures the essential, non-singular features expected from quantum-corrected gravity. Furthermore, by embedding this regularized core within a quintessence field $\sigma$, a physically realistic macroscopic environment compatible with a dark-energy-dominated universe has been constructed. Investigating the extended thermodynamics and Joule-Thomson expansion of this configuration allows us to identify how microscopic regularisations manifest as macroscopic thermodynamic signatures. Ultimately, this approach provides crucial theoretical clues, demonstrating how intrinsic geometric deformations and external exotic matter fields jointly reshape the thermal evolutionary tracks of black holes, offering concrete phase-space behaviours that cannot be obtained in standard GR.}

After the introduction section, we have arranged the rest of the work in this manner: In Section \ref{sec2}, we have discussed the black hole solution of the deformed AdS black hole in the presence of a quintessence field. In Section \ref{section3}, we examine the thermodynamic properties of the black hole and mainly the Joule-Thomson effect of the black hole. Next, the efficiency of the black hole is determined in Section \ref{section4}. Section \ref{sec06} presents our conclusions about our findings for black hole thermodynamics and its efficiency. Throughout this work, we use the natural unit system $G=c=\hbar=1$ and the metric sign convention $(-,+,+,+)$.

\section{Deformed AdS-Schwarzschild black hole in presence of quintessence} \label{sec2}

To obtain the deformed AdS-Schwarzschild black hole solution, we start with the four-dimensional action as \cite{Khosravipoor_2023}
\begin{equation}\label{main action}
    A=\int{d^4 x\,
    \sqrt{-g}\,\left(\frac{R-2\Lambda}{2\kappa}+\mathcal{L}_m+\mathcal{L}_{\rm X}\right)}.
\end{equation}
In this expression, the term $g$ is the determinant of the metric, $R$ is the Ricci scalar, and $\Lambda$ represents the cosmological constant. Also $ \kappa=8\pi$. The Lagrangian density terms, $\mathcal{L}_m$ and $ \mathcal{L}_{\rm X}$ refer to the matter Lagrangian and Lagrangian for theories other than General Relativity (GR) or extra fields like tensor, scalar, or vector fields, respectively. 

Varying this action \eqref{main action} with respect to the metric, one can obtain the field equation as 
\begin{equation}\label{enesteineq}
    G_{\mu \nu} + \Lambda g_{\mu\nu} = \kappa\, T^{(\rm tot)}_{\mu\nu},
\end{equation}
here $G_{\mu\nu}$ is representing the Einstein tensor; $T^{(\rm tot)}_{\mu\nu}$ stands for the total energy-momentum tensor:
\begin{equation}\label{ems}
    T^{(\rm tot)}_{\mu\nu} = T^{(\rm m)}_{\mu\nu} + T^{(\rm X)}_{\mu\nu}.
\end{equation}
$T^{(\rm tot)}_{\mu\nu}$ is the sum of matter and additional fields' energy-momentum tensor, which has the functional forms respectively
\begin{equation}
    T^{(\rm m)}_{\mu\nu}=-\dfrac{2}{\sqrt{-g}}\dfrac{\delta\left(\sqrt{-g}L_m\right)}{\delta g^{\mu\nu}},
\end{equation}
and 
\begin{equation}
    T^{(\rm x)}_{\mu\nu}=-\dfrac{2}{\sqrt{-g}}\dfrac{\delta\left(\sqrt{-g}L_x\right)}{\delta g^{\mu\nu}}.
\end{equation}
For a static and spherically symmetric spacetime, we use the metric function as
\begin{equation}\label{line element1}
    ds^2 = -e^{\nu(r)}dt^2 + e^{\mu(r)}dr^2 + r^2(d\theta^2 + \sin^2{\theta}\,d\phi^2).
\end{equation}
As usual, $t$ and $r$ stand for the time and space coordinates, respectively, and  $\theta$ and $\phi$ are polar and azimuthal angles.

To obtain the required black hole solution, following the article \cite{Khosravipoor_2023}, we shall implement the Kerr-Schild condition \cite{Bini:2010hrs}:
\begin{equation}\label{hcondition}
    e^{\mu (r)}=e^{-\nu (r)}.
\end{equation}
Also, to assimilate the deformation condition, one can introduce an energy density function as \cite{Khosravipoor_2023}:
\begin{equation}\label{Efunction}
    \mathcal{E}(r)=\frac{\alpha }{\kappa  (\beta +r)^4}.
\end{equation}
Here, the term $\beta$ is a constant parameter which controls the behaviour of the energy density at $r=0$, and $\alpha$ stands for the deformation parameter which is responsible for a rapid asymptotic decay, and it avoids a central singularity. { The specific algebraic form of this energy density is strictly motivated by the fundamental physical requirement to resolve the central curvature singularity inherent in the standard GR framework. In this context, the control parameter $\beta$ acts as a necessary short-distance regularisation scale. Conceptually analogous to a minimal length cutoff predicted by effective quantum gravity approaches, it ensures that as the radial coordinate approaches the centre ($r \to 0$), the energy density remains finite rather than diverging to infinity. Simultaneously, the deformation parameter $\alpha$ dictates the strength of this non-singular deviation and enforces a rapid asymptotic decay, ensuring the geometry recovers its standard behaviour at large distances. Consequently, these parameters are not merely arbitrary mathematical artefacts; rather, they serve as physically motivated, effective corrections that capture the macroscopic phenomenological behaviour expected from a regularised, quantum-corrected black hole interior.}

Now using Eq. \ref{Efunction} in the field equation, the metric function $B(r)$ can be obtained as
\begin{equation}\label{Bsolve}
 B(r)=\frac{\frac{\alpha  \beta ^2 }{3 (\beta
+r)^3}+\frac{\alpha } {\beta +r}-\frac{\alpha  \beta }{(\beta
+r)^2}+\frac{r^3+c_1}{l^2}+ r}{ r(1-2 M/r+r^2/l^2)},
\end{equation}
In this expression, $M$ represents the ADM mass, $l=\sqrt{3/|\Lambda|}$ is the AdS radius, and $c_1$ is an integration constant having the value $- 2 M l^2$. From this equation we can have:
\begin{equation}\label{metric fun0}
  e^{\zeta(r)} B(r)=1-\frac{2M}{r}+\frac{r^2}{l^2} +\alpha\,\frac{\beta^2+3r^2+3\beta r}{3r(\beta+r)^3} \equiv F(r).
\end{equation}
Thus, the final line element for the deformed AdS-Schwarzschild black hole reads:
\begin{equation}\label{final line element}
    ds^2 = - F(r) dt^2 + \frac{1}{F(r)} dr^2 + r^2(d\theta^2 + \sin^2{\theta}\, d\phi^2).
\end{equation}
This metric characterises a black hole solution having a deformation parameter $\alpha$ that includes alterations consistent with additional gravitational fields while maintaining the asymptotic AdS behaviour. With this solution, we will now incorporate the deformed Schwarzschild black hole surrounded by a field of quintessence in the AdS background \cite{Shahjalal:2019pqb}. To do so, one can attach the term $(\frac{\sigma }{r^{3\omega_q +1}})$, where $\sigma$ represents a normalization factor associated with the quintessence and $\omega_q$ is a state parameter that satisfies the constraint condition, $\omega_q$ $\in (-1,-1/3)$. 

Thus, the reformed metric function is:
\begin{equation}\label{metric function}
  F(r)=1-\frac{2M}{r}+\frac{r^2}{l^2} +\alpha\,\frac{\beta^2+3r^2+3\beta r}{3r(\beta+r)^3}+\frac{\sigma}{r^{3\omega_q +1}}.
\end{equation} 

In our investigation, we shall use $\omega_q = -2/3$.

In a recent study, a deformed black hole surrounded by both quintessence and a global monopole has been investigated \cite{Ahmed:2025nni}. The above solution is a special case of this solution in the absence of a global monopole.
In the following part of our investigation, we shall use this black hole spacetime to study the thermodynamic behaviour, including the Joule-Thomson effect and thermodynamic efficiency.

\section{Thermodynamics and Joule-Thomson Effect}\label{section3}

The Hawking temperature is a fundamental thermodynamic quantity associated with black hole radiation. It is given by \cite{Bardeen:1973gs}, 
\begin{equation}\label{TH}
    T_H=\frac{\kappa}{2\pi},
\end{equation}
where $\kappa$ is the surface gravity given by,
\begin{equation}\label{kappa}
    \left.\kappa=-\frac{1}{2}\frac{\partial_r g_{tt}}{\sqrt{-g_{tt}g_{rr}}}\right|_{r=r_h}.
\end{equation}
Here $g_{tt}$ is the temporal coefficient and $g_{rr}$ is the spatial coefficient. The term $r_h$ represents the event horizon of the black hole. This results in a finding that the assumption that a black hole is completely black is not true, but it emits thermal radiation, which is known as Hawking radiation.

For the black hole defined by the metric function \eqref{metric function}, the Hawking temperature is calculated as,
\begin{equation}
    T_H = \frac{1}{4} \left(8 P r_h+\frac{2 \sigma -\frac{\alpha  r_h}{\left(\beta +r_h\right){}^4}}{\pi }+\frac{1}{\pi  r_h}\right).
\end{equation}
In the above expression, $P$ is the pressure associated with the black hole spacetime given by,

\begin{equation}
    P = - \dfrac{\Lambda}{8 \pi }.
\end{equation}

We can now use the expressions of Hawking temperature and pressure to obtain the J-T coefficient associated with the black hole spacetime. As mentioned earlier, the J-T expansion is a thermodynamic process which provides us insight into the temperature behaviour of a system undergoing adiabatic expansion. During the J-T expansion, the enthalpy, which is the mass of a black hole, $M$, is constant, making it an isenthalpic process. This allows us to study the temperature changes of the black hole system that result solely from the variations in pressure and volume. The Joule-Thomson coefficient is given by:
\begin{equation}\label{JTcoefficient}
    \mu=\left(\frac{\partial T}{\partial P}\right)_M =\frac{1}{C_p}\left[T\left(\frac{\partial V}{\partial T}\right)_P-V\right]=\frac{(\partial T/\partial r_h)_M}{(\partial P/\partial r_h)_M}
\end{equation}
Here $C_p$ represents the heat capacity at constant pressure of the black hole system. Expression of $C_p$ is given by,
\begin{equation}
    C_p = \frac{2 \pi  r_h \left(8 P r_h+\left(2 \sigma-\frac{\alpha  r_h}{\left(\beta +r_h\right){}^4}+\frac{1}{r_h} \right)\pi^{-1} \right)}{\frac{\alpha  \left(3 r_h-\beta \right)}{\pi  \left(\beta +r_h\right)^5}-\frac{1}{\pi  r_h^2}+8 P}.
\end{equation}

A positive J-T coefficient $(\mu>0)$ indicates cooling upon expansion, while a negative value $(\mu<0)$ corresponds to heating. The following equation shows the slope of the P-T graph for the AdS-deformed black hole surrounded by a quintessence field (with $\omega = -2/3$),

\begin{equation}
    \mu = \frac{2 \pi  r_h^2 \left(8 P- \left(\frac{\alpha }{\left(\beta +r_h\right)^4}-\frac{4 \alpha  r_h}{\left(\beta +r_h\right)^5}+\frac{1}{r_h^2}\right)\pi^{-1} \right) \left(\frac{3 \left(8 P r_h+ \left(-\frac{\alpha  r_h}{\left(\beta +r_h\right)^4}+\frac{1}{r_h}+2 \sigma \right)\pi^{-1} \right)}{8 P- \left(\frac{\alpha }{\left(\beta +r_h\right)^4}-\frac{4 \alpha  r_h}{\left(\beta +r_h\right)^5}+\frac{1}{r_h^2} \right)\pi^{-1} }-r_h\right)}{3 \left(r_h \left(8 \pi  P r_h-\frac{\alpha  r_h}{\left(\beta +r_h\right)^4}+2 \sigma \right)+1\right)}.
\end{equation}

The inversion temperature $T_i$ is defined by the equation $\mu(T_i)=0$. The inversion temperature separates cooling and heating processes. Using the definition of $T_i$ and $\mu$, we have, 
\begin{equation}\label{Ti1}
    T_i=V\left(\frac{\partial T}{\partial V}\right)_P = \frac{r_h}{3}\left(\frac{\partial T}{\partial r_h}\right)_P.
\end{equation}
For the considered deformed AdS-Schwarzschild black hole in the presence of quintessence, the inversion temperature is calculated as,

\begin{equation}
    T_{i} = -\frac{\beta ^5-2 r_h^3 \left(\alpha -5 \beta ^2 (\beta  \sigma +1)\right)+\beta ^4 (\beta  \sigma +5) r_h+5 \beta ^3 (\beta  \sigma +2) r_h^2+(5 \beta  \sigma +1) r_h^5+5 \beta  (2 \beta  \sigma +1) r_h^4+\sigma  r_h^6}{4 \pi  r_h \left(\beta +r_h\right)^5}.
\end{equation}
At inversion pressure $P_i$, $\mu$ equals to zero and therefore $P_i$ has the following form:
\begin{equation}\label{pi}
    P_i= -\frac{2 \beta ^5-\alpha  \beta  r_h^2-3 \alpha  r_h^3+10 \beta ^4 r_h+20 \beta ^3 r_h^2+20 \beta ^2 r_h^3+3 \sigma  r_h \left(\beta +r_h\right){}^5+10 \beta  r_h^4+2 r_h^5}{8 \pi  r_h^2 \left(\beta +r_h\right)^5}.
\end{equation}

\begin{figure}[htbp]
	\centerline{
		\includegraphics[scale = 0.58]{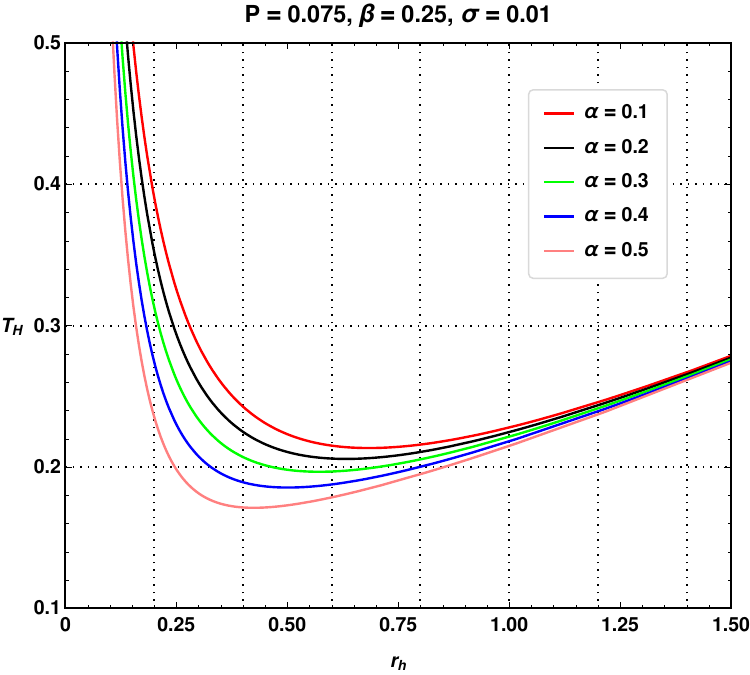}\hspace{0.5cm}
		\includegraphics[scale = 0.58]{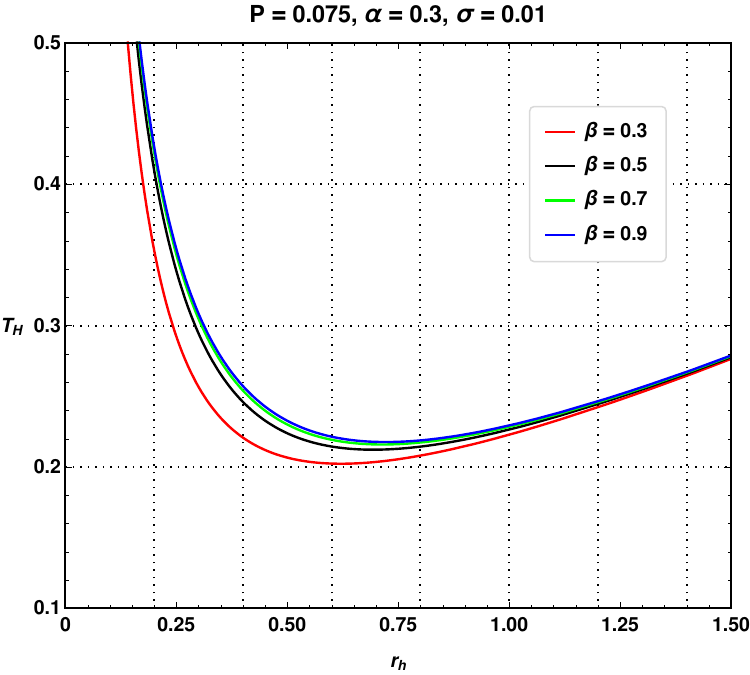}} \vspace{0.5cm}
        \includegraphics[scale = 0.58]{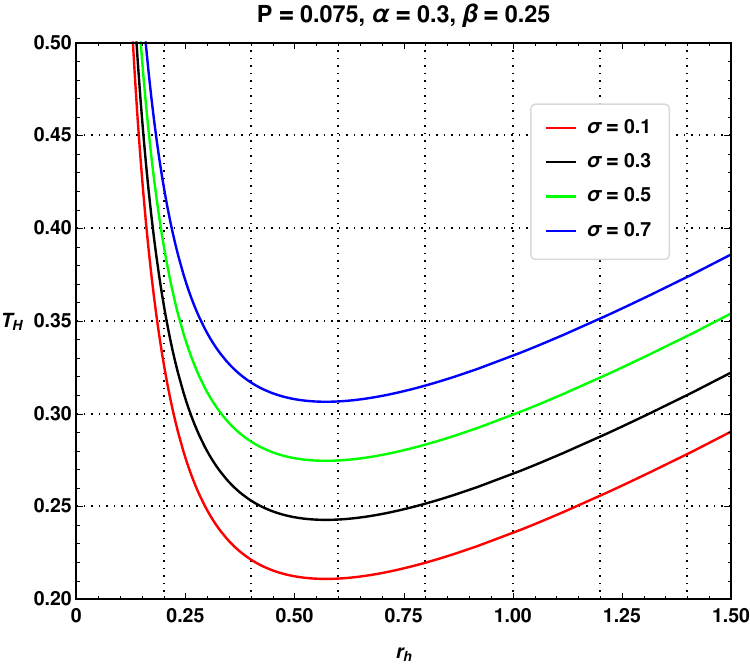}
	\caption{Variation of Hawking temperature with horizon radius of the black hole.}
	\label{Temp}
\end{figure}

Figure~\ref{Temp} illustrates the behaviour of the Hawking temperature $T_{H}$ as a function of the horizon radius $r_{h}$ for different values of the deformation parameter $\alpha$, control parameter $\beta$, and the quintessence parameter $\sigma$. In all cases, the temperature exhibits a characteristic non-monotonic profile: it diverges in the limit of small horizon radius, decreases to a local minimum, and subsequently grows for larger $r_{h}$. This generic structure indicates the presence of a thermodynamic transition between the small and large-horizon phases. The top-left panel displays the influence of $\alpha$, showing that larger values of the deformation parameter lower the minimum temperature and shift its location toward smaller horizon radii. This behaviour reflects the contribution of $\alpha$ to the effective repulsive structure in the near-horizon geometry, thereby altering the stability properties of the black hole.

The influence of the control parameter $\beta$ is shown in the top-right panel. As $\beta$ increases, the temperature curves are slightly elevated, and the temperature minimum becomes shallower and occurs at marginally larger radii. This suggests that $\beta$, which typically encodes nonlinear or regularisation effects, moderates the drop in temperature in the intermediate region, effectively enhancing thermodynamic stability. The modification introduced by $\beta$ becomes more pronounced in the small-radius regime, where nonlinear corrections are strongest, demonstrating its role in controlling the effective gravitational response of the system.

The bottom panel displays the effect of the quintessence parameter $\sigma$. Increasing $\sigma$ results in a higher temperature in both the small and intermediate-radius regimes. The resulting upward shift in $T_{H}$ underscores the thermodynamic sensitivity of the black hole to its external environment. Altogether, the three panels show that the parameters $\alpha$, $\beta$, and $\sigma$ significantly influence the thermal and stability characteristics of the black hole, especially in the region where geometric and matter-field corrections are non-negligible.

\begin{figure}[htbp]
	\centerline{
		\includegraphics[scale = 0.5]{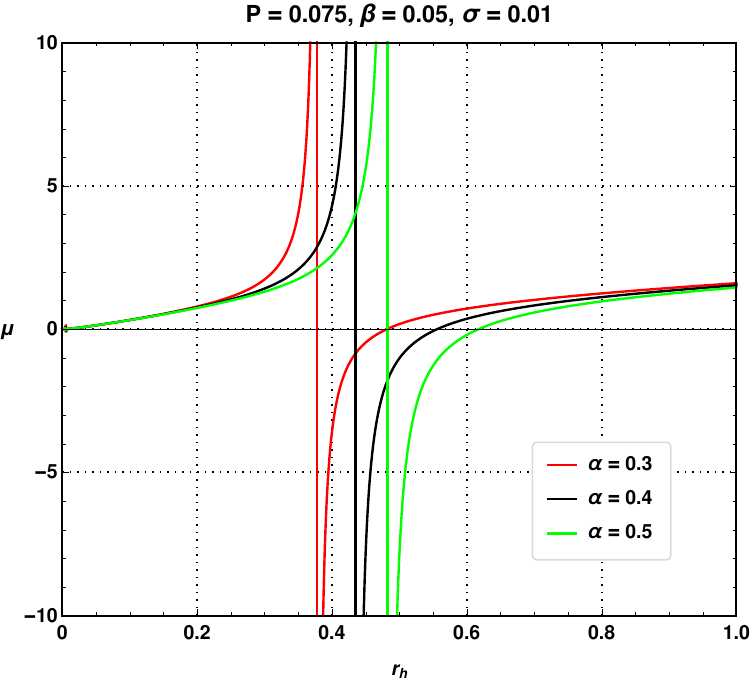}\hspace{0.5cm}
		\includegraphics[scale = 0.5]{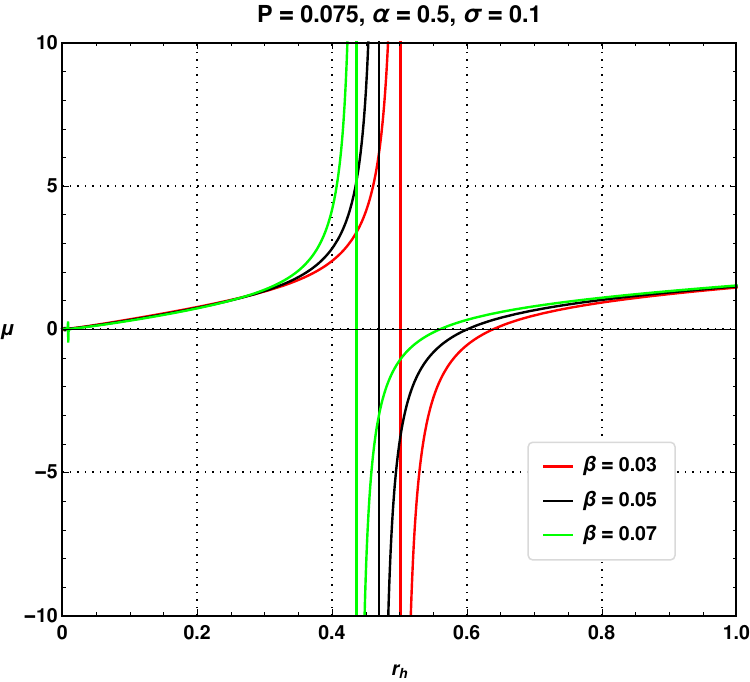}} \vspace{0.5cm}
       \centerline{
		\includegraphics[scale = 0.5]{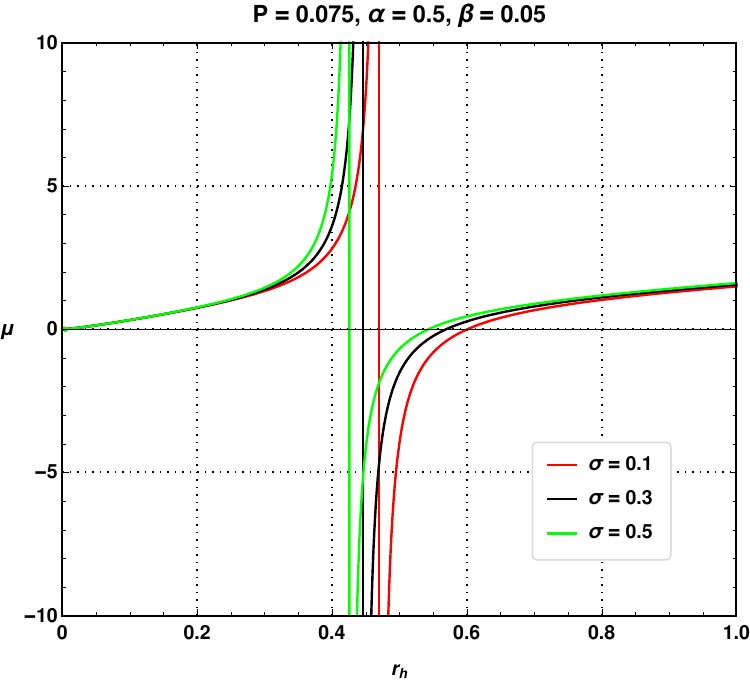}
		}
	\caption{ JT coefficient vs event horizon radius of the black hole.}
	\label{mu01}
\end{figure}

Figure~\ref{mu01} presents the behaviour of the Joule--Thomson coefficient $\mu$ as a function of the event horizon radius $r_{h}$ for different combinations of the deformation parameter $\alpha$, control parameter $\beta$, and quintessence parameter $\sigma$. In each panel, the Joule--Thomson coefficient exhibits a characteristic divergence at a critical radius, corresponding to the point where the slope of the Hawking temperature with respect to pressure changes its sign. The region where $\mu>0$ indicates the cooling phase of the black hole during an isenthalpic expansion, whereas $\mu<0$ corresponds to the heating phase. The presence of both positive and negative branches, therefore, marks the transition from heating to cooling, analogous to the van der Waals gas in ordinary thermodynamics. The sharpness and position of these divergences demonstrate that the black hole response to J-T expansion is highly sensitive to the underlying parameters of the theory.

The first two panels show the influence of $\alpha$ and $\beta$ on $\mu(r_{h})$. Increasing the deformation parameter $\alpha$ shifts the divergence point of $\mu$ toward larger radii and broadens the region over which $\mu$ remains positive, indicating that stronger deformation enhances the cooling regime and delays the transition to heating. Similarly, an increase in $\beta$ produces a comparable shift, though typically with a weaker effect. These trends are in clear agreement with the temperature profiles shown in Fig.~\ref{Temp}, where larger values of $\alpha$ or $\beta$ were found to displace the minimum of the temperature curve to larger horizon radii. Since the inversion point in the Joule--Thomson process coincides with the minimum of the temperature for isenthalpic trajectories, the parallel behaviour of $T_{H}(r_{h})$ and $\mu(r_{h})$ confirms the internal thermodynamic consistency of the model.

The bottom panel highlights the effect of the quintessence parameter $\sigma$. Increasing $\sigma$ causes the divergence of $\mu$ to move to a larger horizon radius, while simultaneously enhancing the magnitude of both the heating and cooling branches. This behaviour directly reflects the upward shift in the Hawking temperature curves observed previously: larger $\sigma$ increases the overall temperature of the black hole at small and intermediate radii, thereby modifying the location of the temperature minimum and the corresponding inversion radius. The consistent correlation between the temperature curves and the Joule--Thomson coefficient demonstrates that the parameters $\alpha$, $\beta$, and $\sigma$ collectively reshape the thermal landscape of the black hole. Together, the figures confirm that both the cooling–heating transition and the inversion mechanism are tightly governed by the same structural features of the black-hole geometry and its surrounding matter fields.

\begin{figure}[htbp]
	\centerline{
		\includegraphics[scale = 0.45]{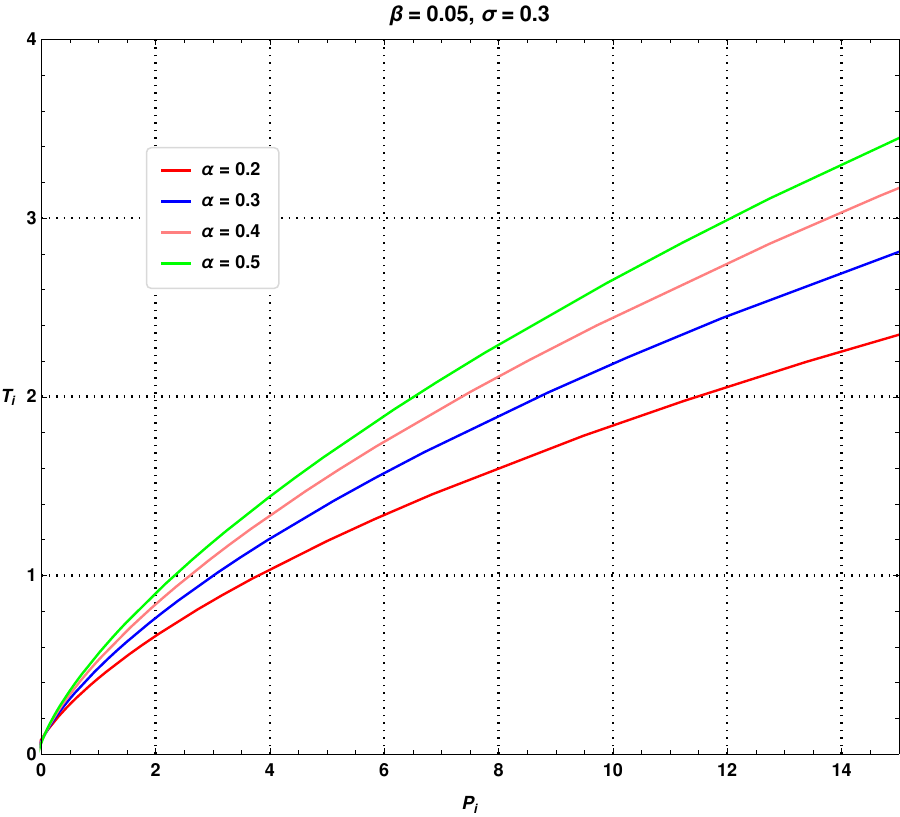}\hspace{0.5cm}
		\includegraphics[scale = 0.45]{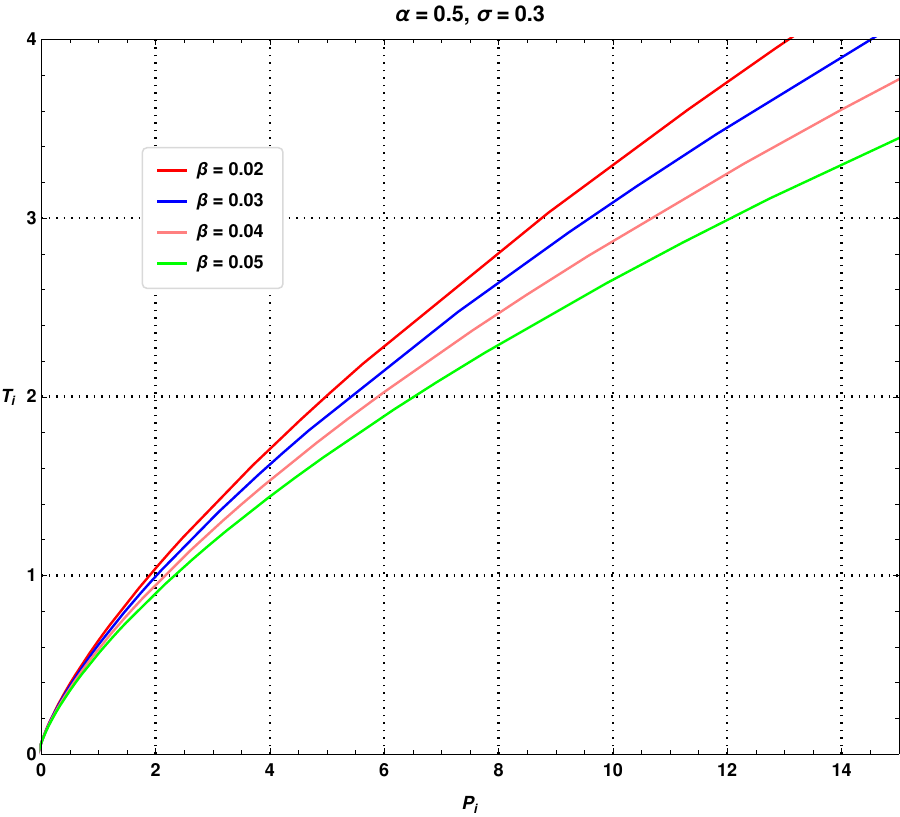}} \vspace{0.5cm}
       \centerline{
		\includegraphics[scale = 0.45]{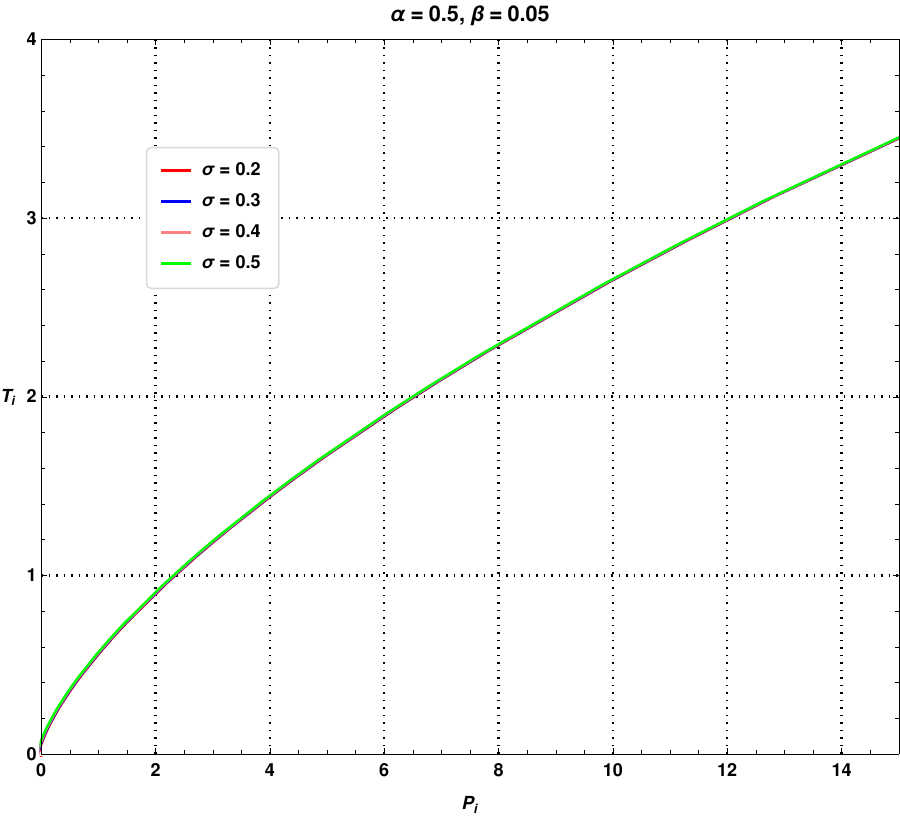}\hspace{0.5cm}
		\includegraphics[scale = 0.45]{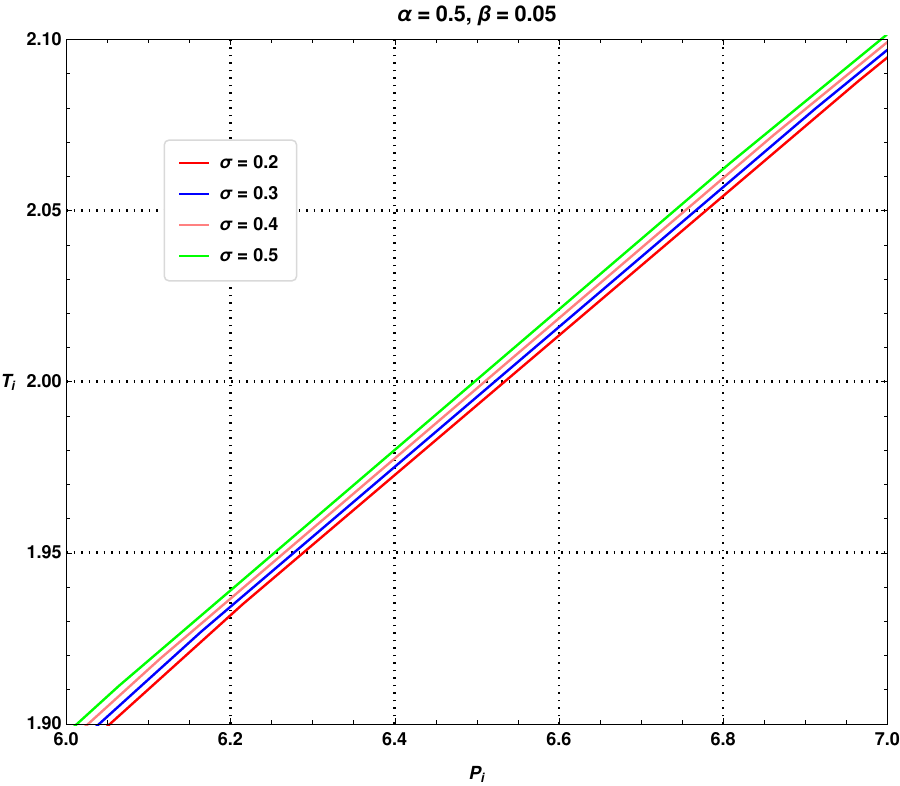}}
	\caption{ Inversion curves of the black hole.}
	\label{inv01}
\end{figure}

Figure~\ref{inv01} displays the inversion curves of the black hole for different values of the parameters $\alpha$, $\beta$, and $\sigma$. The inversion temperature $T_{i}$ marks the boundary between heating $(\mu<0)$ and cooling $(\mu>0)$ during the J-T expansion, and therefore encodes crucial information about the thermodynamic response of the system. In all panels, the inversion temperature increases monotonically with the inversion pressure $P_{i}$, reflecting the characteristic behaviour familiar from van der Waals fluids. The first figure demonstrates the influence of $\alpha$, showing that larger values of the deformation parameter shift the inversion curve upward. This indicates that a stronger deformation increases the temperature at which the cooling–heating transition occurs. This trend is consistent with the behaviour observed in both the temperature and J-T coefficient plots: increasing $\alpha$ moves the temperature minimum to a larger radius and enhances the cooling region, thereby requiring a higher inversion temperature to trigger the transition.

The effect of the control parameter $\beta$ is illustrated in the second panel. Similar to $\alpha$, increasing $\beta$ raises the inversion temperature across the full pressure range, although the shift is comparatively weaker. Since $\beta$ typically controls the strength of nonlinear or regularisation effects in the underlying geometry, its influence on the inversion curve reflects modifications to the repulsive contributions in the thermodynamic equation of state. These results reinforce the trends previously observed in Fig.~\ref{Temp} and Fig.~\ref{mu01}: larger $\beta$ flattens the temperature profile, shifts the critical radius associated with the temperature minimum, and moderately alters the divergence structure of $\mu(r_{h})$. The corresponding upward shift in $T_{i}$ is therefore a natural thermodynamic consequence of the interplay between $\beta$ and the near-horizon modifications introduced by the matter sector.

The final two panels show the effect of the quintessence parameter $\sigma$. The third figure reveals a slight increase in the inversion temperature as $\sigma$ grows, while the fourth figure provides a zoomed-in view confirming that the influence of $\sigma$ on the inversion curve is positive but extremely small. This behaviour is fully consistent with the earlier findings: in Fig.~\ref{Temp}, increasing $\sigma$ raised the Hawking temperature, particularly in the small- and intermediate-radius regions, while Fig.~\ref{mu01} demonstrated that $\sigma$ shifts the divergence of the Joule--Thomson coefficient toward larger radii. Since the inversion temperature is determined by the condition $\mu=0$, these changes naturally translate to a mild upward shift in $T_{i}$. However, the comparatively weaker dependence on $\sigma$ indicates that the external quintessence-like field does not significantly modify the compressibility or microstructure of the black hole, leading to only minimal changes in its expansion-driven thermodynamic behaviour.

\begin{figure}[htbp]
	\centerline{
		\includegraphics[scale = 0.45]{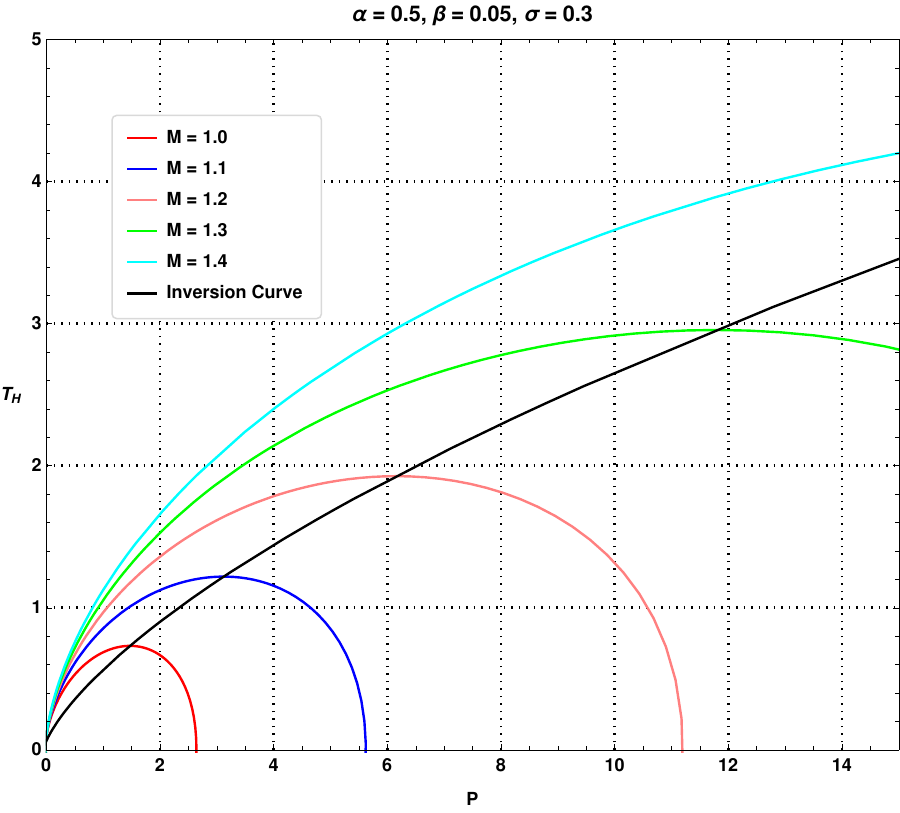}\hspace{0.5cm}
		\includegraphics[scale = 0.45]{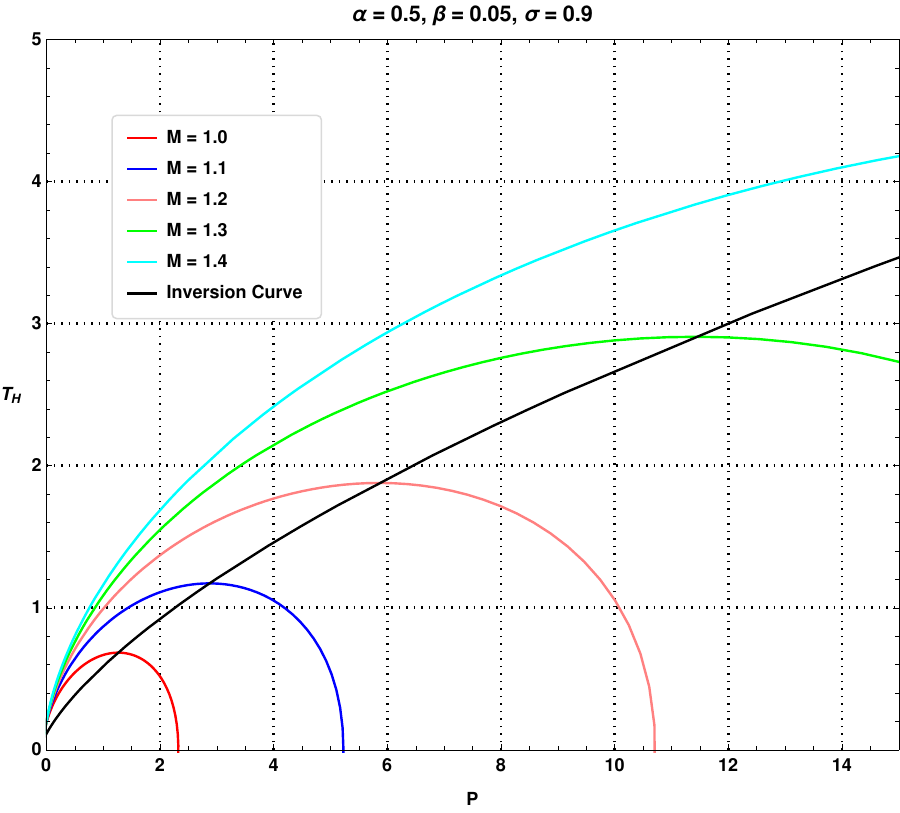}} \vspace{0.5cm}
       \centerline{
		\includegraphics[scale = 0.45]{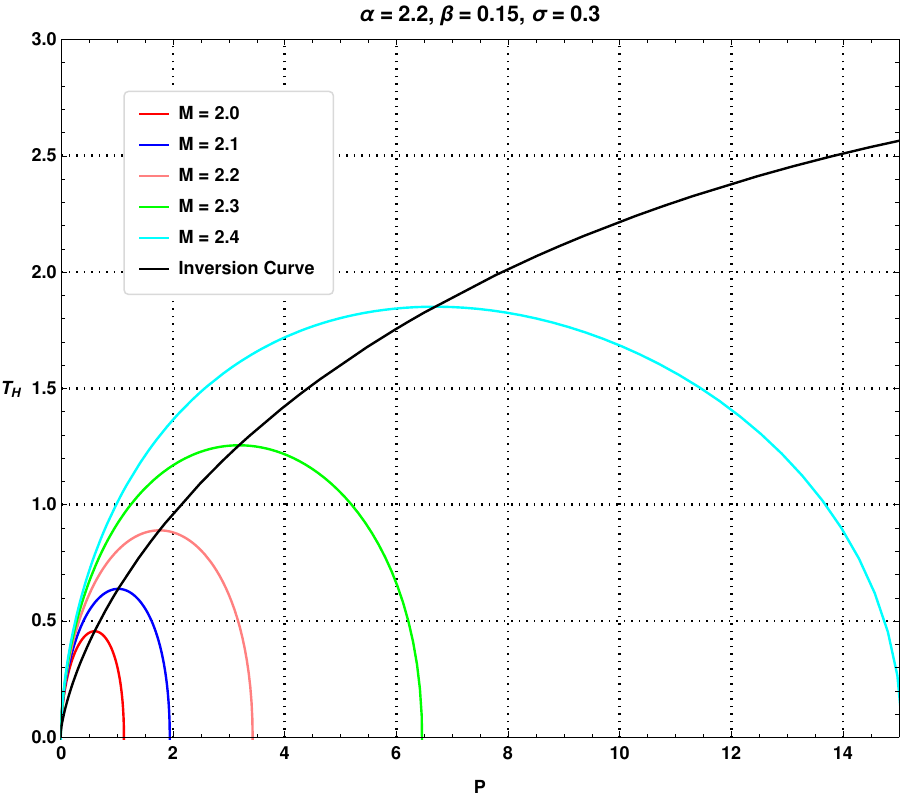}\hspace{0.5cm}
		\includegraphics[scale = 0.45]{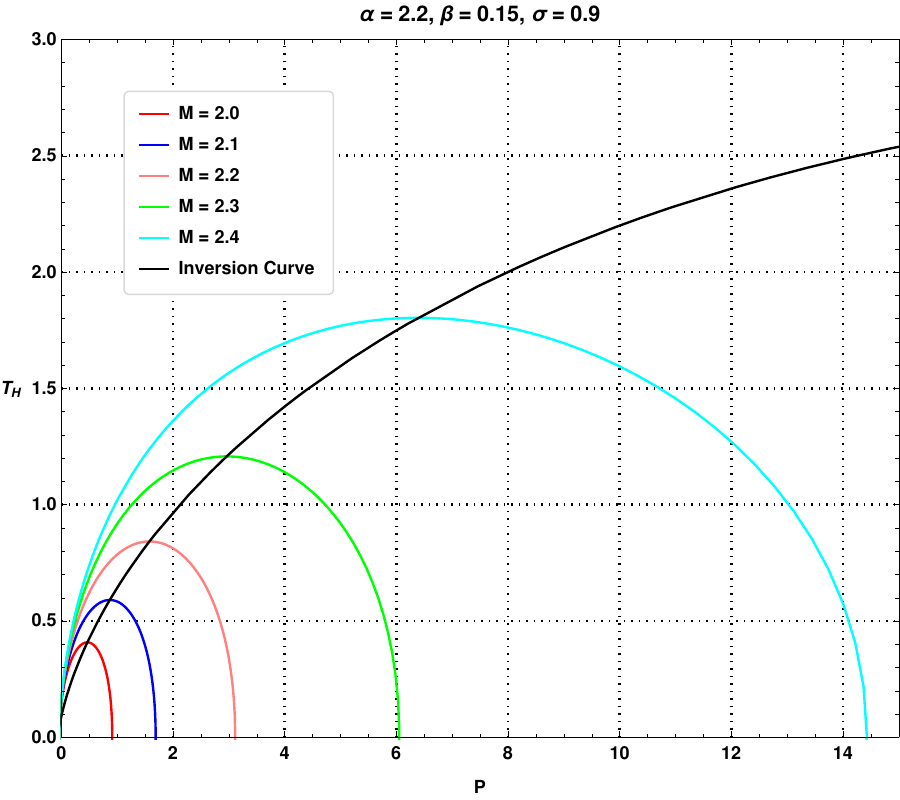}}
	\caption{ Isenthalpic and inversion curves of the black hole.}
	\label{isen01}
\end{figure}

%In Figures \ref{isen01} and \ref{isen02}, we have shown the isenthalpic and inversion curves for the black hole for different values of the model parameters. 

%In Figure \ref{isen01}, the first two plots show the behaviour of the insenthalpic curves for $\sigma=0.3$ and $\sigma=0.9$ with $\alpha=0.5$ and $\beta=0.05$. The third and fourth plots show the behaviour of the isenthalpic curves for similar values of $\sigma$ but with $\alpha=2.2$ and $\beta=0.15$. These plots suggest that the impact of the quintessence parameter $\sigma$ is amplified by an increase in the parameters $\alpha$ and $\beta$.

\begin{figure}[htbp]
	\centerline{
		\includegraphics[scale = 0.45]{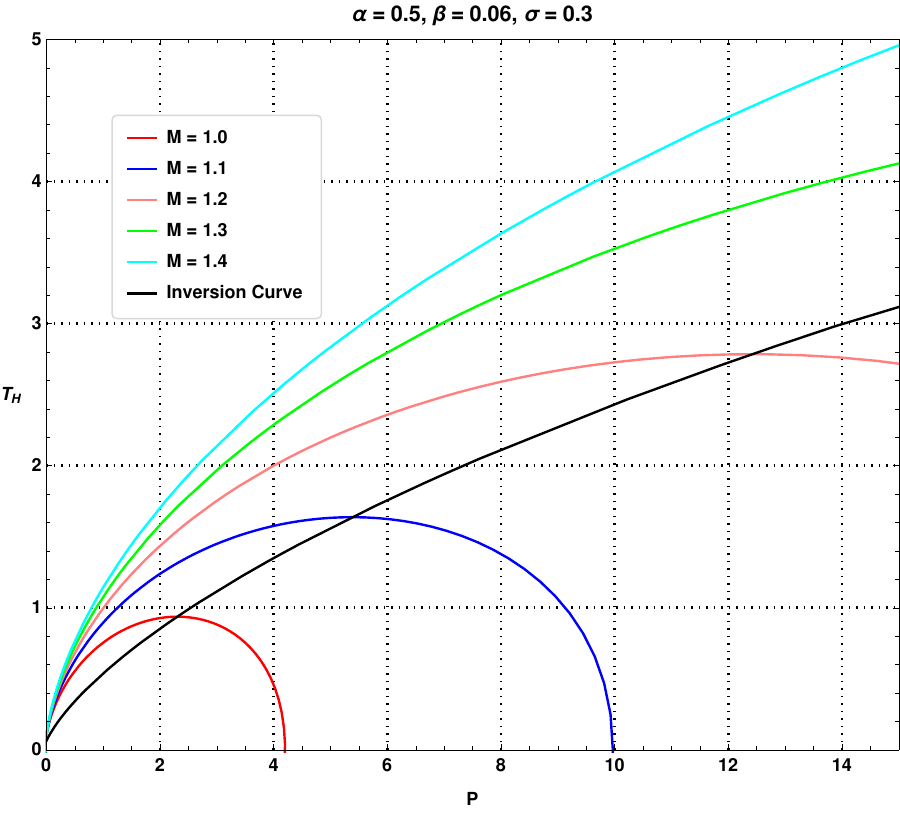}\hspace{0.5cm}
		\includegraphics[scale = 0.45]{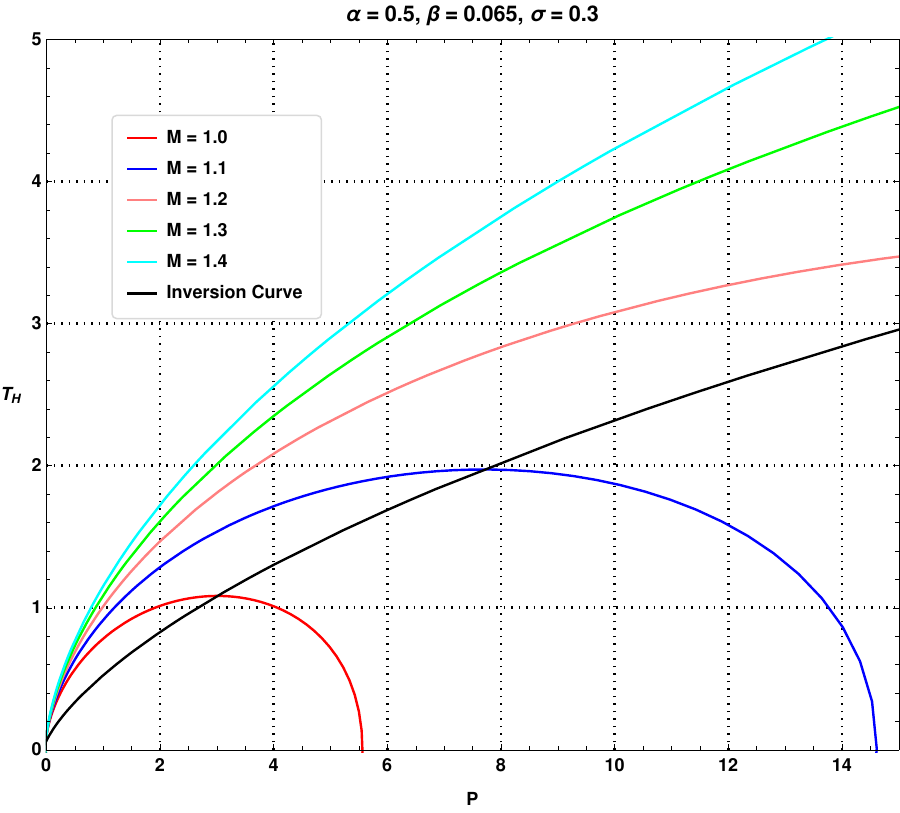}} \vspace{0.5cm}
       \centerline{
		\includegraphics[scale = 0.45]{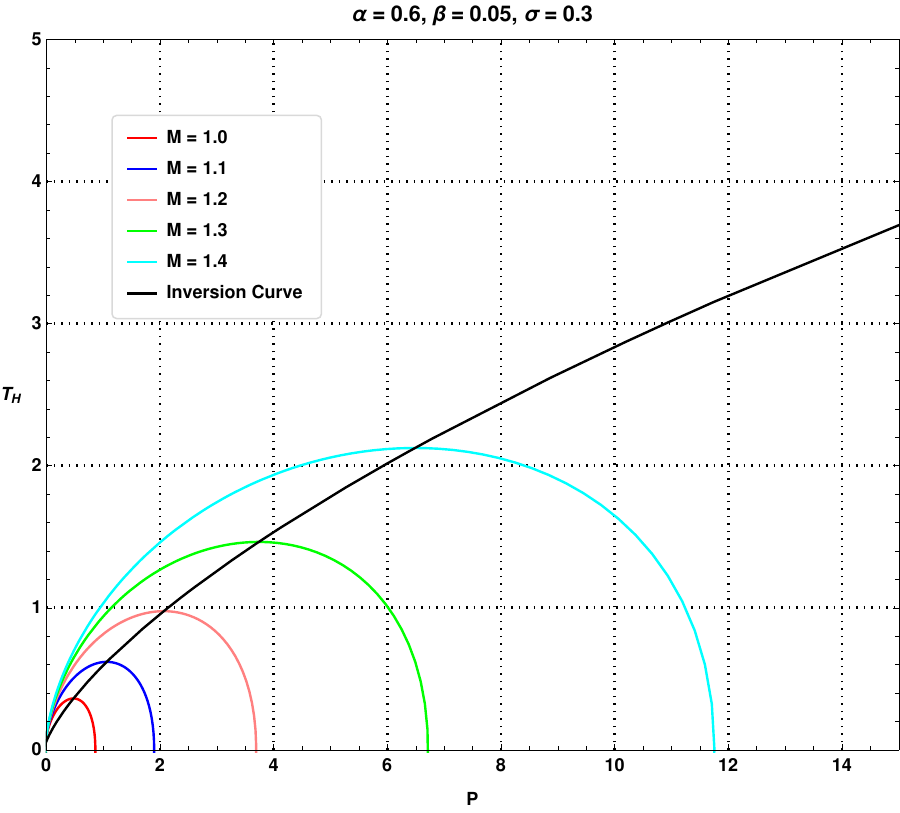}\hspace{0.5cm}
		\includegraphics[scale = 0.45]{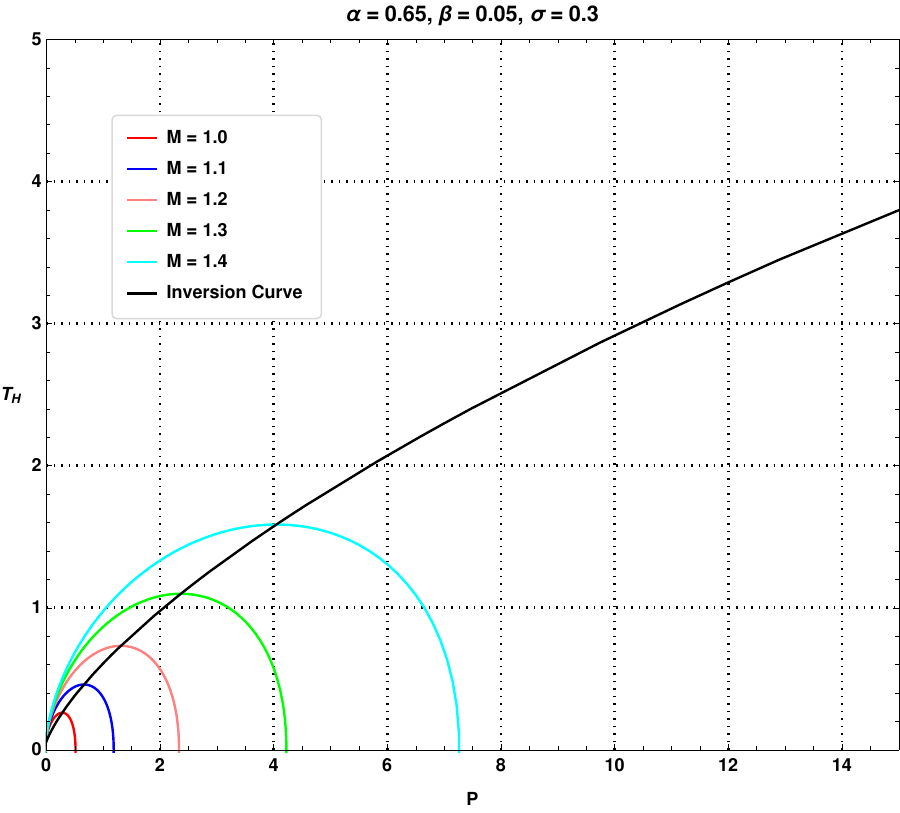}}
	\caption{ Isenthalpic and inversion curves of the black hole.}
	\label{isen02}
\end{figure}

Figures~\ref{isen01} and~\ref{isen02} illustrate the isenthalpic trajectories of the black hole in the $(P,T)$ plane for various values of the model parameters, together with the corresponding inversion curves. For a fixed enthalpy $M$, each isenthalpic curve exhibits a characteristic behaviour: the temperature initially increases with pressure, reaches a maximum, and then decreases before intersecting the inversion curve. The point of intersection marks the transition between the heating $(\mu<0)$ and cooling $(\mu>0)$ phases during a Joule--Thomson expansion. Curves lying above the inversion line correspond to regimes in which the black hole cools under expansion, whereas those lying below indicate heating. The position and curvature of these isenthalpic trajectories therefore, encode detailed information about how the internal structure of the black hole responds to changes in the external thermodynamic variables.

The first two panels of Fig.~\ref{isen01} show the isenthalpic curves for $\alpha=0.5$ and $\beta=0.05$ for two different values of the quintessence parameter, $\sigma=0.3$ and $\sigma=0.9$. Increasing $\sigma$ shifts the isenthalpic curves upward and slightly modifies the location of the turning points, although the overall shape remains similar. This behaviour is consistent with the earlier temperature analysis in Fig.~\ref{Temp}, where larger $\sigma$ was found to elevate the Hawking temperature, particularly in the small- and intermediate-radius regimes. The mild upward shift of the isenthalpic curves reflects the modest influence of the surrounding quintessence fluid on the thermal response of the black hole, in agreement with the small but positive shift observed in the inversion temperature in Fig.~\ref{inv01}. Thus, for fixed values of $\alpha$ and $\beta$, the quintessence parameter affects the cooling–heating transition but only weakly, indicating that its thermodynamic effect is secondary to that of the intrinsic geometric parameters.

A more pronounced behaviour appears in the third and fourth panels, corresponding to $\alpha=2.2$ and $\beta=0.15$. In this regime, the isenthalpic curves display a stronger sensitivity to $\sigma$: the trajectories shift upward more substantially, and the departure between curves corresponding to different $\sigma$ becomes more visible across the full pressure range. This amplification of the $\sigma$-dependence is consistent with the trends observed in earlier sections. Larger values of $\alpha$ and $\beta$ were shown to enhance the repulsive or regularising effects in the near-horizon region, thereby modifying both the temperature minima and the divergence structure of the J-T coefficient. As a result, the external quintessence field interacts more strongly with the underlying geometry, producing a more significant shift in the cooling and heating regions. The combined behaviour confirms that the thermodynamic influence of $\sigma$ becomes increasingly important when the deformation and nonlinear parameters are enhanced, demonstrating a nontrivial interplay between the intrinsic structure of the black hole and the surrounding exotic matter distribution.

%In Figure \ref{isen02}, first and second plots are produced with $\beta = 0.06$ and $\beta=0.065$ which suggest that an increase in the control parameter $\beta$ increases the area under the isenthalpic curves.

%Third and fourth plots are produced using $\alpha = 0.6$ and $\alpha=0.65$ which suggest that an increase in the value of the deformation parameter $\alpha$ decreases the area under the isenthalpic curve.

Figure~\ref{isen02} further illustrates the behaviour of the isenthalpic curves in the $(P,T)$ plane, focusing on the influence of the control parameter $\beta$ and the deformation parameter $\alpha$. The first two panels correspond to $\beta=0.06$ and $\beta=0.065$ with fixed $\alpha=0.5$ and $\sigma=0.3$. A clear trend emerges: increasing $\beta$ causes the isenthalpic trajectories to expand outward, thereby increasing the area enclosed beneath each curve. Physically, this expansion reflects the enhanced nonlinear contributions associated with larger $\beta$, which modify the effective equation of state of the black hole. This behaviour is consistent with earlier findings from the temperature plots and the inversion curves, where larger $\beta$ elevated the temperature and shifted the inversion point. The increased area under the isenthalpic curves therefore confirms that the thermodynamic response of the black hole becomes more sensitive to pressure changes when nonlinear effects are strengthened.

The third and fourth panels show the isenthalpic curves for $\alpha=0.6$ and $\alpha=0.65$, keeping $\beta=0.05$ and $\sigma=0.3$ fixed. In contrast to the effect of $\beta$, increasing the deformation parameter $\alpha$ leads to a reduction in the area under the corresponding isenthalpic curves. This contraction indicates that higher values of $\alpha$ suppress the thermal response of the black hole at fixed enthalpy, a behaviour aligned with the shift of the temperature minimum and the earlier observation that $\alpha$ enhances the cooling region in the Joule--Thomson process. Since larger $\alpha$ introduces stronger geometric or regularisation effects near the horizon, the corresponding isenthalpic trajectories become more constrained, reducing the extent of the thermodynamic excursion in the $(P,T)$ plane.

Together, the results in Fig.~\ref{isen02} reinforce the complementary roles played by $\alpha$ and $\beta$ in the black-hole thermodynamics. While $\beta$ amplifies the nonlinear matter-sector contributions and broadens the isenthalpic curves, $\alpha$ strengthens the geometric deformation and correspondingly contracts the accessible region of the $(P,T)$ phase space. These trends are fully consistent with the behaviour observed in Figs.~\ref{Temp}, \ref{mu01}, and \ref{inv01}, demonstrating a coherent interplay between the model parameters and the cooling--heating structure of the black hole.

\section{Efficiency of the Black Hole as a Thermal System}\label{section4}

\begin{figure}[h!]
    \centering
    \includegraphics[height=8.5cm, width=9.5cm]{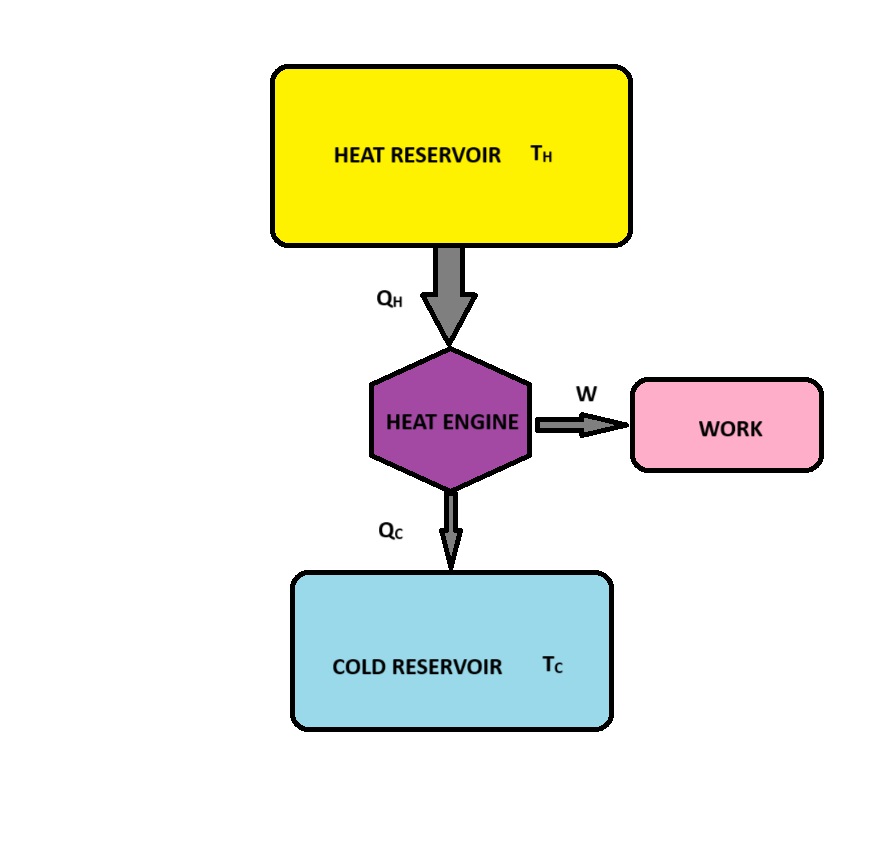}
    \includegraphics[height=7cm, width=8cm]{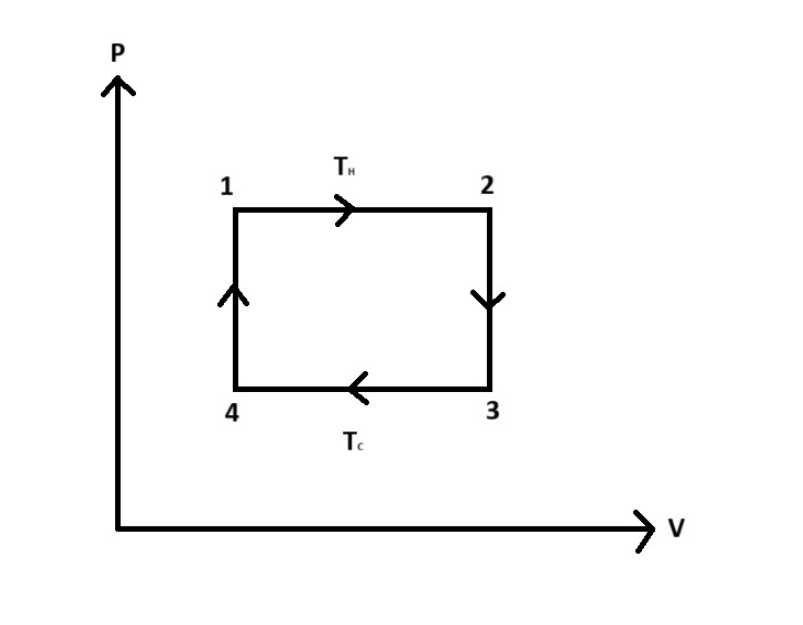}
    \caption{A schematic figure of the working of a heat engine is shown. On the right, we have the Carnot engine P-V cycle.}
    \label{figR1}
\end{figure}

A heat engine is any device or mechanism that converts heat into useful mechanical work. It gets energy from a source, which sits at a higher temperature, and after some part of this heat absorbed is converted to useful work, the rest amount of heat is transferred to the sink, which sits at a lower temperature. There is a working substance, usually a gas, expanding and contracting between some cycles of change, which facilitates the functioning of the system. More explicitly, it involves a closed-path process in the P-V plane involving absorption of heat $Q_H$ from the source and eventually eliminating $Q_C$ amount of heat to the sink. The work done is $W=Q_H-Q_C$. From the definition of efficiency of a heat engine, we have \cite{Rajani:2019ovp}
\begin{equation}
    \eta= \frac{W}{Q_H}.
\end{equation}
The maximum efficiency possible for a heat engine is theoretically the Carnot efficiency, given by \cite{Rajani:2019ovp}:
\begin{equation}
    \eta_c = 1-\frac{Q_C}{Q_H}=1-\frac{T_C}{T_H}.
\end{equation}
Here, $T_C$ and $T_H$ represent the sink and source temperatures, respectively. In the case of static black hole system \cite{Rajani:2019ovp}, it involves a simple reversible cycle with isotherms at higher and lower temperatures. Isothermal expansion involves the absorption of $Q_H$ heat, and during the isothermal compression, $Q_C$ heat is released. These paths are connected by adiabatic paths as can be seen from figure \ref{figR1}. For the sake of avoiding complications, we adopt a rectangular P-V cycle ($1\to 2\to 3\to 4\to 1$) as shown in figure \ref{figR1}. 
The amount of work done in one complete operating cycle of the black hole heat engine is mathematically stated as:
\begin{equation}
    W_{net}=W_{1\to 2}+W_{2\to3}+W_{3\to4}+W_{4\to1}=\frac{4}{3\sqrt{\pi}} (P_{1}-P_{4})\Big(S_2^{\frac{3}{2}}-S_1^{\frac{3}{2}}\Big).
\end{equation}
Here, the terms $P_1$ and $P_4$ stand for pressure at states 1 and 4 of P-V cycle, respectively. Similarly, $S_2$ and $S_1$ represent the entropy of the system at states 2 and 1. As no heat exchange takes place in the isochoric processes, we compute heat absorbed during the isothermal expansion of the system $Q_H$, (during $1\to2$) as:
\begin{equation}
  Q_H=\int_{T_1}^{T_2} C_P (P_1,T)dT=\int_{S_1}^{S_2}C_P\Big(\frac{\partial T}{\partial S}\Big)dS=\int_{S_1}^{S_2}TdS=M_2-M_1.  
\end{equation}

\begin{equation}
Q_H= \frac{1}{6}\Bigg[\frac{3\big(\sqrt{S_{2}}-\sqrt{S_{1}}\big)+8P_{1}\big(S_{2}^{3/2}-S_{1}^{3/2}\big)}{\sqrt{\pi}}
+\alpha\!\left(
\frac{\dfrac{3S_{2}}{\pi}+\dfrac{3\sqrt{S_{2}}\beta}{\sqrt{\pi}}+\beta^{2}}
{\left(\dfrac{\sqrt{S_{2}}}{\sqrt{\pi}}+\beta\right)^{3}}
-\frac{\dfrac{3S_{1}}{\pi}+\dfrac{3\sqrt{S_{1}}\beta}{\sqrt{\pi}}+\beta^{2}}
{\left(\dfrac{\sqrt{S_{1}}}{\sqrt{\pi}}+\beta\right)^{3}}
\right)
+\frac{3\sigma\,(S_{2}-S_{1})}{\pi}
\Bigg]
\end{equation}
The efficiency of the black hole heat engine is expressed as:
\begin{multline}
\eta= \frac{8 \left(P_1-P_4\right) \left(S_2^{3/2}-S_1^{3/2}\right)}{\sqrt{\pi } \left(-\frac{8 P_1 S_1^{3/2}}{\sqrt{\pi }}+\frac{8 P_1 S_2^{3/2}}{\sqrt{\pi }}-\frac{\alpha  \left(\beta ^2+\frac{3 \beta  \sqrt{S_1}}{\sqrt{\pi }}+\frac{3 S_1}{\pi }\right)}{\left(\beta +\frac{\sqrt{S_1}}{\sqrt{\pi }}\right){}^3}+\frac{\alpha  \left(\beta ^2+\frac{3 \beta  \sqrt{S_2}}{\sqrt{\pi }}+\frac{3 S_2}{\pi }\right)}{\left(\beta +\frac{\sqrt{S_2}}{\sqrt{\pi }}\right){}^3}-\frac{3 \sigma  S_1}{\pi }+\frac{3 \sigma  S_2}{\pi }+\frac{3 \sqrt{S_2}}{\sqrt{\pi }}-\frac{3 \sqrt{S_1}}{\sqrt{\pi }}\right)}
\end{multline}

The following figure \ref{figR2} shows the variation of efficiency versus entropy $S_2$ for different values of deformation parameter $\alpha$. It is seen that with increasing $\alpha$, efficiency shows a fair bit of variation for smaller $S_2$. After that, the curves converge, and efficiency becomes constant at about 90\%. Lower $\alpha$ implies lower efficiency. The next plot shows the variation of efficiency with $\beta$. We observed that efficiency is larger for smaller values of the control parameter $\beta$, before converging to 88\% for all the graphs. The third plot shows the variation of efficiency with quintessence parameter $\sigma$. It is seen that efficiency rises to about 80 \% for very smaller values of $S_2$, then decreases for all cases. It also converges to about 52\% efficiency with increasing $S_2$. Lower $\sigma$ leads to higher efficiency. It is worth mentioning that though we do not encounter negative efficiency, this may imply a reverse heat engine cycle, behaving like a refrigerator.   

\begin{figure}[h!]
    \centering
    \includegraphics[scale = 0.5]{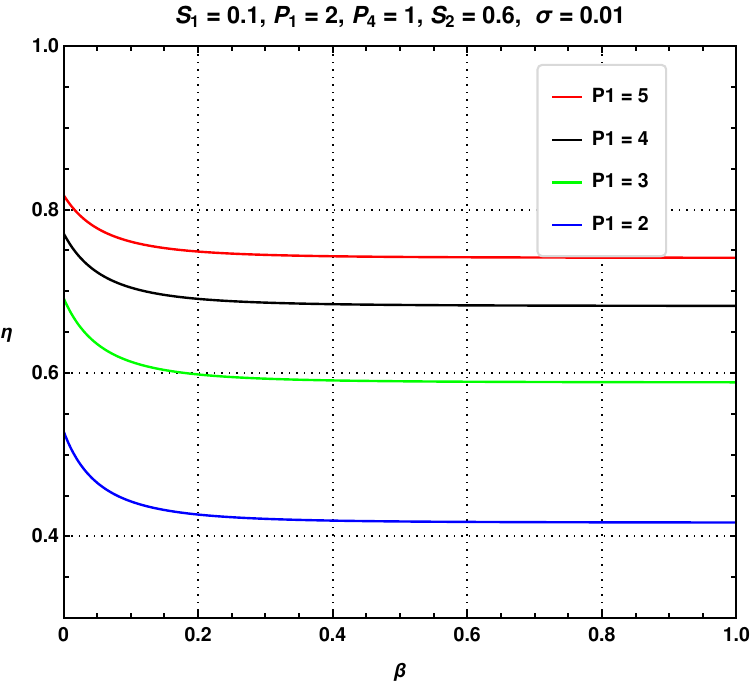}\hspace{0.4cm}
    \includegraphics[scale = 0.5]{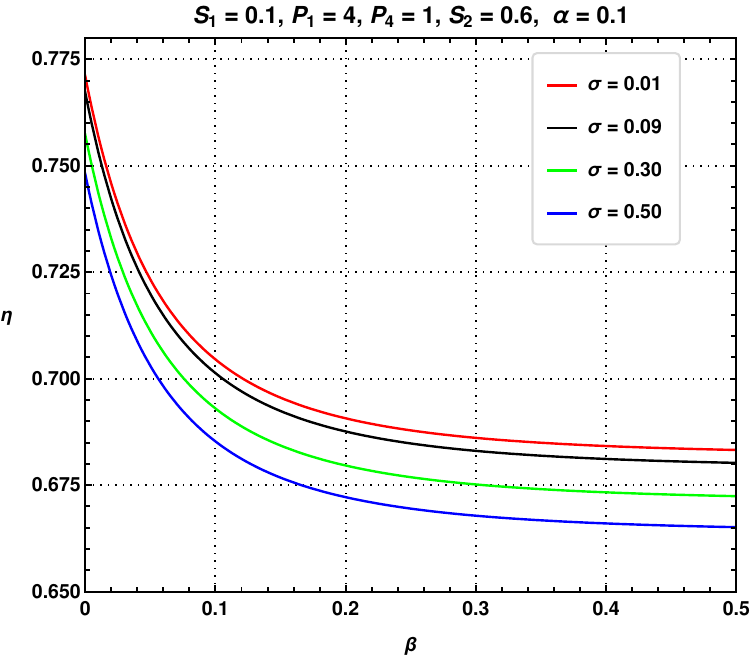}
    \caption{The efficiency of a deformed AdS-Schwarzschild black hole in the presence of
quintessence field is shown in the plots. The variation of efficiency with $\beta$ is analysed with different values of $P_1$ and pressure term $\sigma$ is shown.}
    \label{figR4}
\end{figure}

In the next Figure \ref{figR3}, variation of efficiency with deformation parameter $\alpha$ is analysed with various values of the control parameter $\beta$ and pressure $P_1$ respectively. As can be seen from the left plot, the efficiency increases more rapidly for smaller $\beta$ values. Efficiency increases with increasing $P_1$ as seen from the graph. Similarly, the Figure \ref{figR4} shows variation of efficiency versus $\beta$ for various values of pressure $P_1$ and quintessence parameter $\sigma$ respectively. Efficiency decreases with $\beta$ and it is also seen from the first plot that efficiency is higher for larger pressure. Similarly, efficiency is larger for smaller quintessence parameter $\sigma$ as can be clearly inferred from the plot.   
\begin{figure}[h!]
    \centering
    \includegraphics[scale = 0.45]{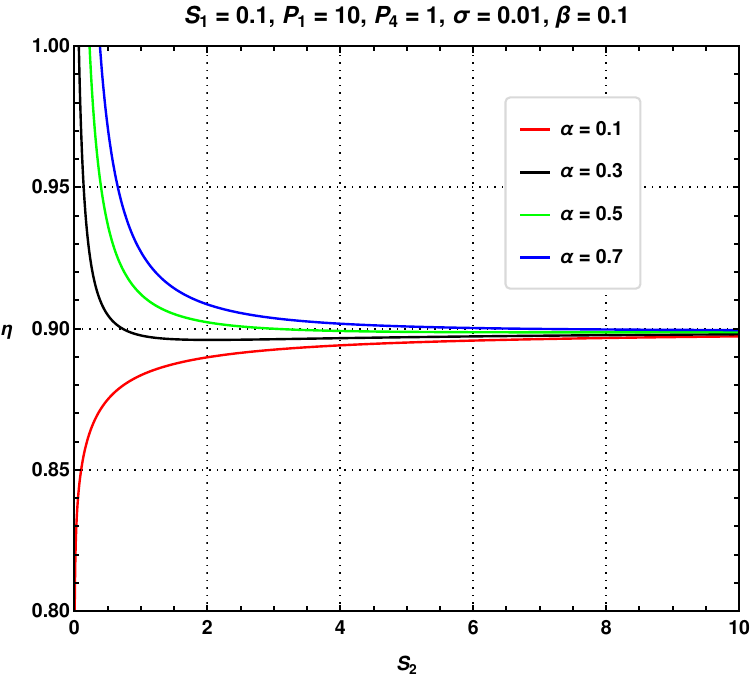}\hspace{0.2cm}
    \includegraphics[scale = 0.45]{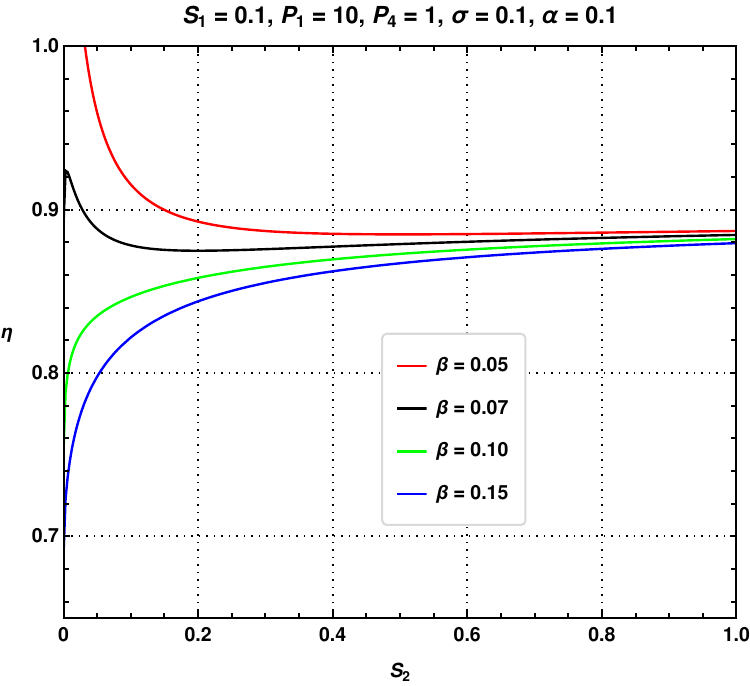}\hspace{0.2cm}
    \includegraphics[scale = 0.45]{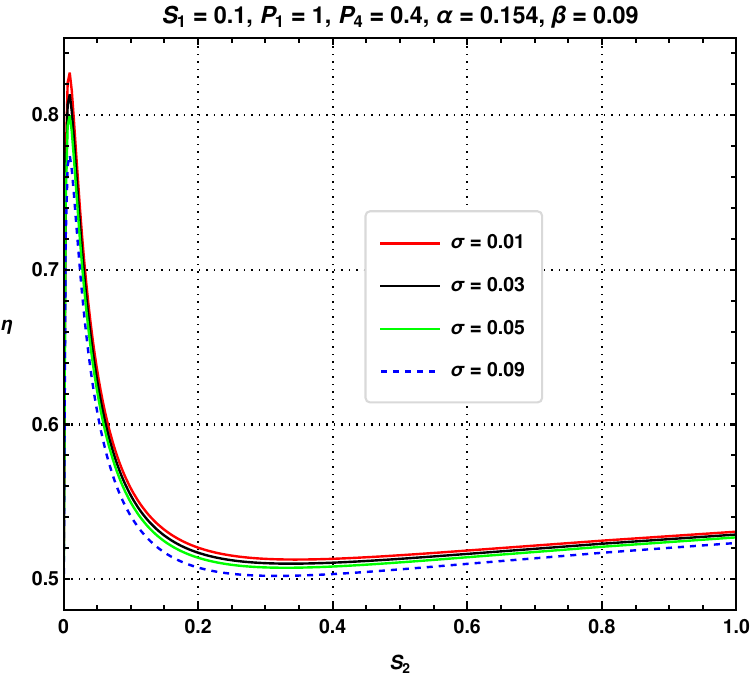}
    \caption{The efficiency of a deformed AdS-Schwarzschild black hole in the presence of
quintessence field is shown in the plots. We analyse the variation of efficiency with respect to $S_2$ for different values of deformation parameters $\alpha$ and $\beta$, and quintessence parameter $\sigma$ in the plots.}
    \label{figR2}
\end{figure}
\begin{figure}[h!]
    \centering
    \includegraphics[scale = 0.5]{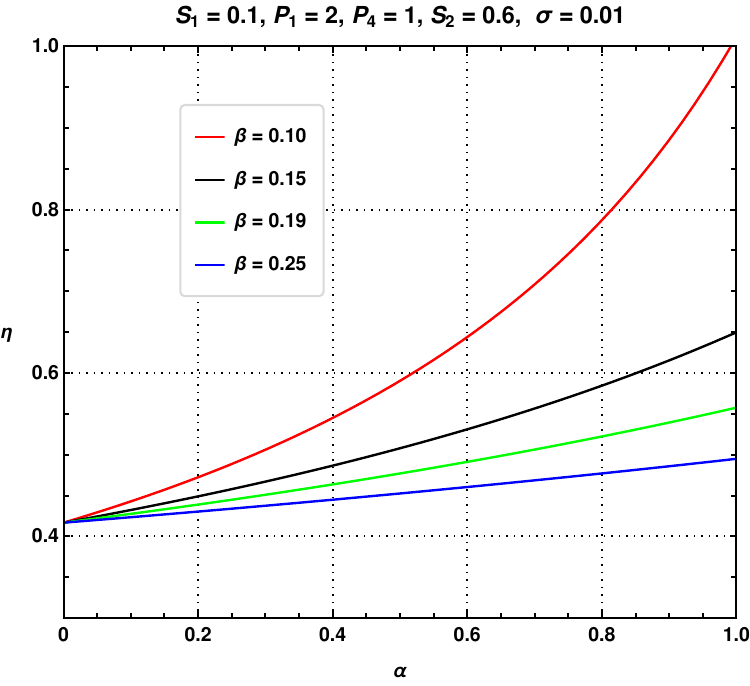}\hspace{0.4cm}
    \includegraphics[scale = 0.5]{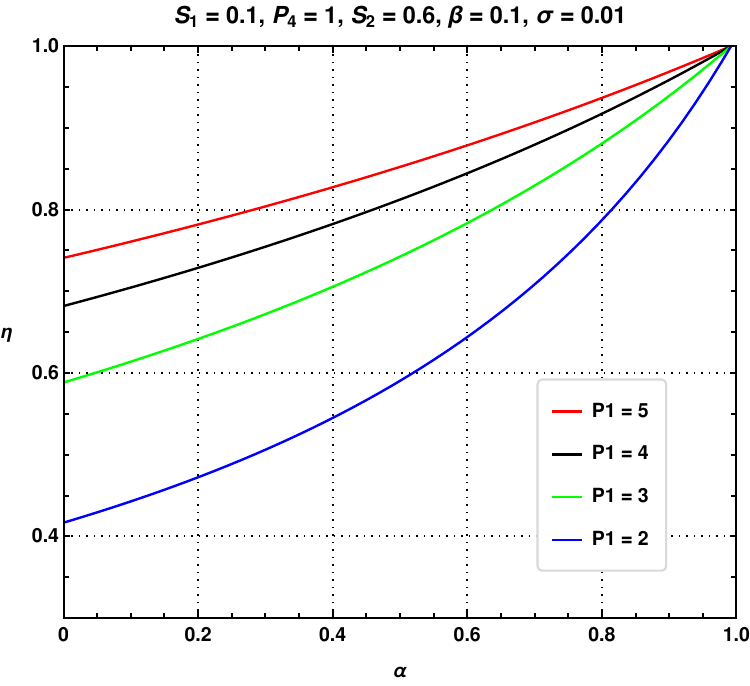}
    \caption{The efficiency of a deformed AdS-Schwarzschild black hole in the presence of
quintessence field is shown in the plot. The variation of efficiency with $\alpha$ is analysed with different values of $\beta$ and pressure term $P_1$ is shown.}
    \label{figR3}
\end{figure}
\begin{figure}
    \centering
    \includegraphics[width=0.3\linewidth]{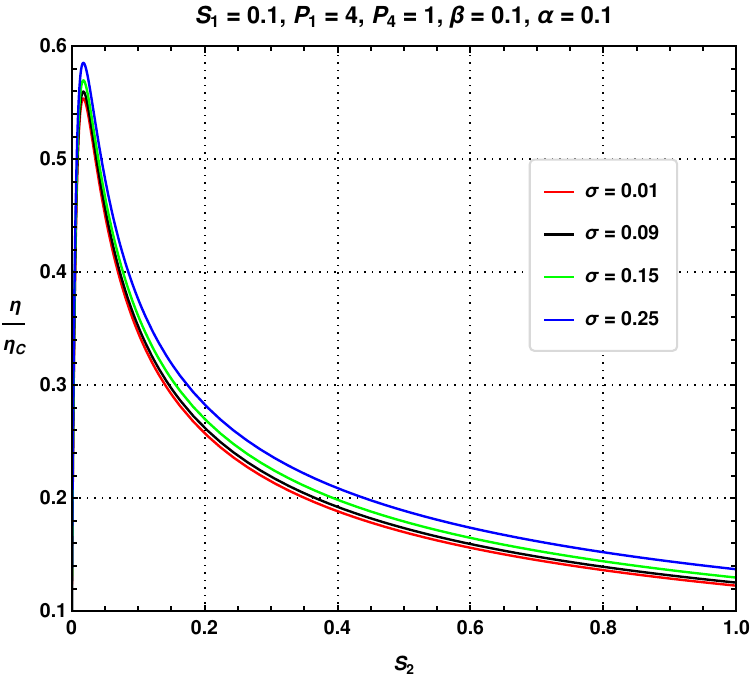}
    \includegraphics[width=0.3\linewidth]{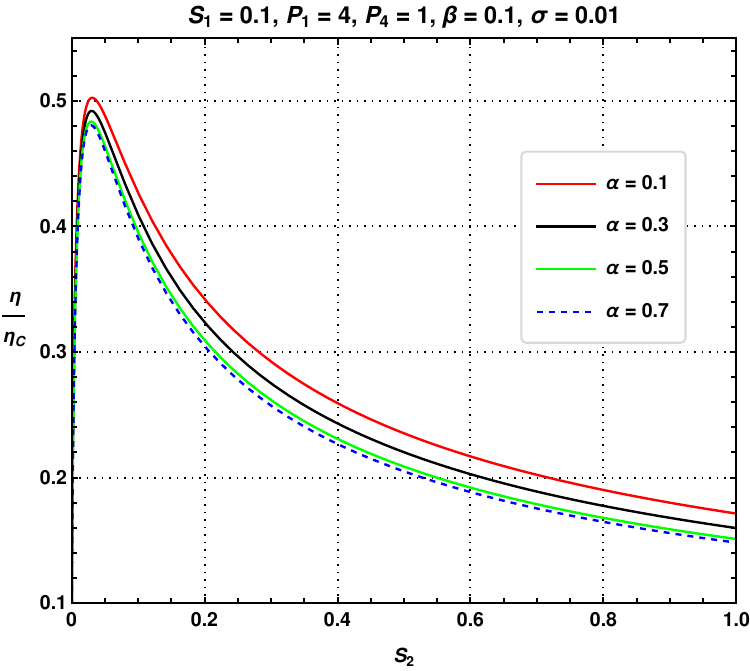}
    \includegraphics[width=0.3\linewidth]{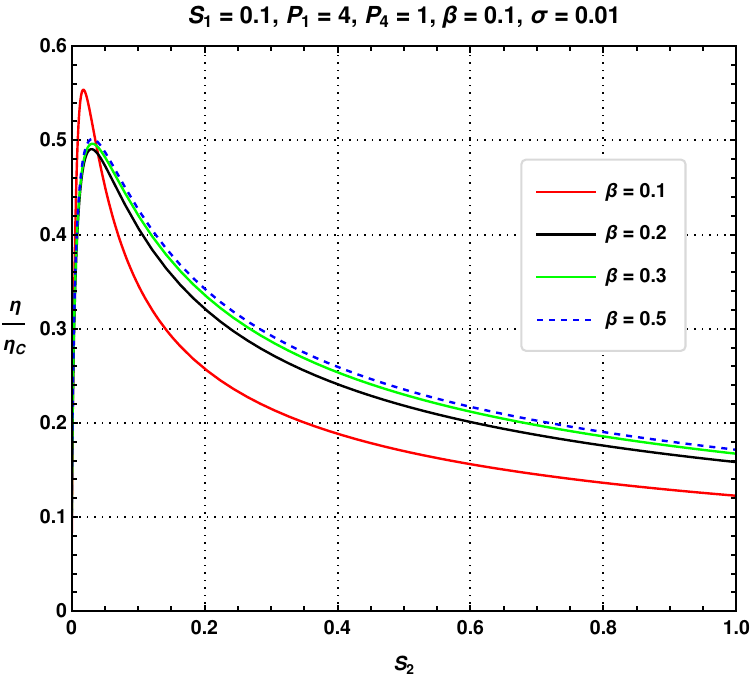}
    \caption{Ratio of efficiency to Carnot efficiency of a deformed AdS-Schwarzschild black hole in the presence of quintessence field is shown in the plots. Variation with $S_2$ is displayed for different values of $\sigma$, $\alpha$ and $\beta$ respectively.}
    \label{figR5}
\end{figure}

The expression for Carnot efficiency is found out to be:
\begin{equation}
    \eta_c= 1-\frac{T_C}{T_H}=1 - \frac{\frac{\pi  \left(8 P_4 S_1+1\right)}{\sqrt{S_1}}+2 \sqrt{\pi } \sigma -\frac{\alpha  \sqrt{S_1}}{\left(\beta +\frac{\sqrt{S_1}}{\sqrt{\pi }}\right){}^4}}{\frac{\pi  \left(8 P_1 S_2+1\right)}{\sqrt{S_2}}+2 \sqrt{\pi } \sigma -\frac{\alpha  \sqrt{S_2}}{\left(\beta +\frac{\sqrt{S_2}}{\sqrt{\pi }}\right){}^4}}
\end{equation}
The ratio of efficiency to Carnot efficiency is plotted for different values of quintessence parameter $\sigma$, deformation parameters $\alpha$ and control parameter $\beta$ respectively, as shown in the figure \ref{figR5}. The steps of computation can be followed from Ref. \cite{Rajani:2019ovp}. The first plot displays efficiency ratio $\frac{\eta}{\eta_c}$ versus $S_2$ for different values of quintessence parameter $\sigma$. The efficiency ratio increases as $\sigma$ increases. The second plot displays the efficiency ratio versus entropy $S_2$ for various values of $\alpha$. The efficiency ratio increases as $\alpha$ decreases. Third plot is efficiency ratio versus $S_2$ for different control parameter values. Except for $\beta=0.1$, higher $\beta$ values signify higher efficiency ratio. The values of the parameters chosen have also been mentioned in the plots. It is also important to note that the chosen values of the parameters are for the proper display of the results and further study has to be done to constrain the exact range of the parameters. In this analysis, we can see some tentative ranges of parameters beyond which efficiency is going towards negative range. This may indicate that the black hole heat engine might not be feasible in that regime of parameter values. 

\begin{table}[h!]
\centering
\renewcommand{\arraystretch}{1.25}
\setlength{\tabcolsep}{10pt}
\caption{{Numerical values of efficiency and the ratio of efficiency to Carnot efficiency for the black hole heat engine.}
Parameters used: $P_{4}=1$, $S_{2}=0.6$, $S_{1}=0.1$}
\begin{tabular}{|c c c c c c|}
\toprule
 \textbf{$P_1$} & \textbf{$\alpha$} & \textbf{$\beta$} & \textbf{$\sigma$} & \textbf{$\eta/\eta_c$} & \textbf{$\eta$} \\
\hline
 2.0 & 0.1 & 0.1 & 0.01 & $0.3006$ & $0.4428$ \\
 3.0 & 0.1 & 0.1 & 0.01 & $0.2056$  & $0.6138$ \\
 2.0 & 0.3 & 0.1 & 0.01 & $0.0529$  & $0.5061$ \\
 2.0 & 0.3 & 0.3 & 0.01 & $0.3813$  & $0.4312$ \\
 2.0 & 0.1 & 0.1 & 0.05 & $0.3030$  & $0.4409$ \\
\hline
\end{tabular}
\label{tabR1}
\end{table}

The tabular presentation of $\eta/\eta_c$ and $\eta$ for some particular values of deformation parameters $\alpha$, control parameter $\beta$ and quintessence parameter $\sigma$ has been shown in Table \ref{tabR1}. The values of efficiency (ratio) and its dependence on the model parameters is shown, which is clearly visible from Figure \ref{figR5}. Parameter values are chosen such that the efficiency remains rational and positive.

%\section{Results}\label{section5}
\section{Concluding Remarks}\label{sec06}

In this work, we have investigated the J-T expansion and the extended thermodynamic behaviour of a black hole modified by the deformation parameter $\alpha$, control parameter $\beta$, and quintessence parameter $\sigma$. By analysing the Hawking temperature, the J-T coefficient, the inversion curves, and the associated isenthalpic trajectories, we obtained a comprehensive picture of how these parameters influence both the microscopic structure and macroscopic thermal response of the system.

The temperature analysis revealed that all three parameters significantly affect the location and depth of the temperature minimum, thereby modifying the thermal stability of small black holes. An increase in $\alpha$ or $\beta$ shifts the temperature minimum to larger horizon radii, whereas $\sigma$ predominantly elevates the temperature in the small-radius regime. Correspondingly, the behaviour of the J-T coefficient exhibits divergences whose locations and signs are in agreement with these temperature trends: larger values of $\alpha$ and $\beta$ enlarge the cooling region, while $\sigma$ produces a milder but consistent shift in the heating–cooling transition. These observations confirm that the cooling and heating characteristics of the black hole are tightly linked to the interplay between geometric deformation, nonlinear matter coupling, and external exotic fluid.

The inversion curves further solidify this thermodynamic picture. Both $\alpha$ and $\beta$ raise the inversion temperature across the full pressure range, indicating that the black hole requires higher temperatures to undergo the transition between heating and cooling when the deformation or nonlinear effects are stronger. Although the influence of $\sigma$ on the inversion temperature is comparatively weak, it remains positive and consistent with its effect on the Hawking temperature and the J-T coefficient. The isenthalpic analysis also reveals clear trends: increasing $\beta$ expands the isenthalpic trajectories, while increasing $\alpha$ compresses them, demonstrating opposite but complementary roles played by the two intrinsic parameters in shaping the thermodynamic phase space.

The black hole heat engine is theorised, and the efficiency is analysed for variation of model parameters. The deformation parameter $\alpha$ increases the efficiency, while a higher $\beta$ value leads to lower efficiency of the heat engine. Heat engine efficiency has an inverse relation with quintessence parameter $\sigma$. 

{ By comparing our results with past related works, several fundamental physical consequences emerge that distinguish this deformed configuration from both standard General Relativity and other regular black hole models. Most notably, our heat engine analysis reveals a contrasting behavior: while past studies, such as those on the Bardeen-AdS black hole \cite{Rajani:2019ovp}, demonstrate that a quintessence field enhances thermodynamic efficiency, our deformed model shows the exact opposite. Specifically, both the quintessence parameter $\sigma$ and the control parameter $\beta$ actively reduce the heat engine's efficiency. Furthermore, our model exhibits a unique competing thermodynamic interplay in the $(P,T)$ phase space that cannot be captured by standard GR. The deformation parameter $\alpha$ contracts the accessible area under the isenthalpic curves, thereby suppressing the thermal response, whereas $\beta$ amplifies nonlinear matter-sector contributions, causing these trajectories to expand outward. Ultimately, these geometric and matter-field corrections act collectively to govern the inversion mechanism, forcing the black hole to require higher temperatures to trigger the heating-cooling transition. This unified thermodynamic structure underscores the profound differences between our specific deformed configuration and previously studied regular black holes.}

Taken together, these results demonstrate that the thermodynamic and J-T properties of the black hole are highly sensitive to modifications introduced by $\alpha$, $\beta$, and $\sigma$. The coherent agreement between the temperature profiles, J-T coefficient behaviour, inversion curves, and isenthalpic trajectories provides strong evidence that the extended phase structure is governed by a unified underlying mechanism shaped by geometric and matter-field corrections. This study not only deepens our understanding of modified black-hole thermodynamics but also opens avenues for exploring analogous phenomena in other regular, nonlinear, or exotic-matter-supported gravitational systems. Future work may involve analysing critical behaviour, microstructure interactions, or quantum corrections within this framework, potentially revealing further connections between modified gravity and black-hole heat-engine physics.

\section*{Acknowledgments}
DJG acknowledges the contribution of the COST Action CA21136  -- ``Addressing observational tensions in cosmology with systematics and fundamental physics (CosmoVerse)". RK is thankful to Prof. U. D. Goswami for his mentorship during RK's early days of research.

\section*{Data Availability Statement}
There are no new data associated with this article.

\bibliography{ref}

%\end{thebibliography}
\end{document}